# Self-Organized Criticality in the Brain

**Dietmar Plenz*, Tiago L. Ribeiro, Stephanie R. Miller, Patrick A. Kells, Ali Vakili, Elliott L. Capek**

Section on Critical Brain Dynamics, National Institute of Mental Health, National Institutes of Health, Bethesda, MD, USA

*** Correspondence:** Dr. Dietmar Plenz, plenzd@mail.nih.gov

## Abstract

Self-organized criticality (SOC) refers to the ability of complex systems to evolve towards a 2nd-order phase transition at which interactions between system components lead to scale-invariant events beneficial for system performance. For the last two decades, considerable experimental evidence accumulated that the mammalian cortex with its diversity in cell types and connections might exhibit SOC. Here we review experimental findings of isolated, layered cortex preparations to self-organize towards four dynamical motifs identified in the cortex *in vivo*: up-states, oscillations, neuronal avalanches, and coherence potentials. During up-states, the synchronization observed for nested theta/gamma-oscillations embeds scale-invariant neuronal avalanches that exhibit robust power law scaling in size with a slope of -3/2 and a critical branching parameter of 1. This dynamical coordination, tracked in the local field potential (nLFP) and pyramidal neuron activity using 2-photon imaging, emerges autonomously in superficial layers of organotypic cortex cultures and acute cortex slices, is homeostatically regulated, displays separation of time scales, and reveals unique size vs. quiet time dependencies. A threshold operation identifies coherence potentials; avalanches that in addition maintain the precise time course of propagated synchrony. Avalanches emerge under conditions of external driving. Control parameters are established by the balance of excitation and inhibition (E/I) and the neuromodulator dopamine. This rich dynamical repertoire is not observed in dissociated cortex cultures, which lack cortical layers and exhibit dynamics similar to a 1st-order phase transition. The precise interactions between up-states, nested oscillations, avalanches, and coherence potentials in superficial cortical layers provide compelling evidence for SOC in the brain.

## Introduction

Brains are inherently complex. Composed of a vast number of cell types, orders of magnitudes larger number of connections, and a myriad of structural and functional networks that make up biochemical pathways affecting every spatial and temporal scale of brain organization – brains are deeply challenging to study. Yet, elaborate efforts to assemble the rich and detailed structural evidence on brain circuits have uncovered a rather small set of dynamical features. Highly detailed brain models comprised of thousands of neurons exhibit relatively simple neuronal activity patterns that range from irregular firing to synchronized or oscillatory activity similar to what is measured in real brains (Markram et al., 2015; Dura-Bernal et al., 2019). Importantly, the major aspect of brain dynamics that has been particularly difficult to understand thus far, is how many neurons in cortex selectively communicate over long distances with associated characteristic times and level of coherence.

The aspect of many interacting elements leading to relatively few dynamical motifs is also a major appeal of self-organized criticality (SOC) (Bak et al., 1987). SOC will drive a system towards a second-order phase transition at which dynamics are dominated by universal properties (for review see (Jensen, 1998; Chialvo, 2010; Mora and Bialek, 2011; Pruessner, 2012; Hesse and Gross, 2014; Marković and Gros, 2014; Muñoz, 2018)). The universal property that is of particular interest to brain functions is scale-invariance indicative of system-wide correlations that emerge in a system exhibiting SOC. Such scale-invariance could be a hallmark of coordinated, yet adaptive, neuronal activity that incorporates large numbers of brain cells.

For the brain, and specifically the cortex, it is currently an intensive field of research whether certain aspects of brain dynamics are true aspects of SOC. Fortunately, numerous key features of SOC can be addressed experimentally in a number of advanced brain preparations (Plenz, 2012). For example, one would expect cortical tissue, developing autonomously in isolation, i.e. in the absence of any instructive sensory and motor inputs, to exhibit scale-invariant properties in the emergent dynamics. One would expect that the emergence of scale-invariance is highly regulated as well. For example, it should be robust to slow driving (i.e., exhibiting a separation of time scales); it should exhibit homeostatic regulation (i.e., returning to scale-invariance after profound perturbations), with these regulations failing when essential circuit components are absent or suppressed.

This review summarizes experimental findings on the emergent dynamics of immature and mature cortical networks when taken in isolation, and thus disconnected from any external, structuring input or required outputs. These dynamics from cortical tissue in isolation, i.e. *in vitro*, will be compared to corresponding dynamical findings in the intact brain, i.e. *in vivo*. It will be argued that the four dynamical motifs of up-states, nested oscillations, neuronal avalanches, and coherence potentials emerge in superficial layers of cortex as major hallmarks of SOC in the brain.

## STRUCTURAL MOTIFS OF SELF-ORGANIZATION: CORTICAL LAYERS, PYRAMIDAL NEURONS, INTERNEURONS, AND GLIAL CELLS

Until now, the organotypic cortex culture to date represents the most complex *in vitro* model of the cortex. Typically taken from a newborn rodent and grown in isolation for up to several months (Fig. 1), it captures several core features of cortical organization. First, it exhibits the major division of the mammalian cortex into superficial and deep cortical layers (Fig. 1A, B; (Götz and Bolz, 1992; Plenz and Aertsen, 1996b; a; Gorba et al., 1999), which exhibit distinct functional properties (Luhmann et al., 2016; Molnár et al., 2020). Superficial layers 2/3, called the associative layers, are composed of pyramidal (excitatory) neurons with reduced branching of their apical dendrites that preferentially connect to other intralaminar pyramidal neurons (Fig. 1C, *top*). In contrast, pyramidal neurons from deep layer 5/6 typically feature elaborate apical dendrites and, besides selectively connecting with superficial layers, communicate *in vivo* with brain regions outside cortex (Fig. 1C, *bottom*; (Plenz and Kitai, 1998)). *In vivo*, layer 4 receives sensory input via the thalamus, a brain structure that conveys sensory information to the cortex; this selective connectivity has been found as well for organotypic co-cultures using thalamus and cortex (Bolz et al., 1990; Gähwiler et al., 1997; Humpel, 2015). The second important hallmark in cortical organization is the presence of three major interneuron (inhibitory) classes identified as parvalbumin (PV), somatostatin (SST) and vasoactive intestinal peptide (VIP) expressing neurons, which exhibit highly selective connectivity and specific firing patterns (for review see (Tremblay et al., 2016; Lim et al., 2018)). Several of these classes, with their layer-specific distribution and electrophysiology, have been demonstrated in organotypic cortex cultures using various immunochemical markers (Fig. 1B; (Götz and Bolz, 1989; Plenz and Aertsen, 1996a; Klostermann and Wahle, 1999)). The third and often overlooked hallmark of the cortical microcircuit is the up to 10x higher presence of non-neuronal cells, or glial cells, compared to neurons. Of the three types of glial cells, cortical astrocytes exhibit brain region specific control over neuronal excitation and dynamics, amongst many other functions (Halassa et al., 2007; Fellin et al., 2009; Perea et al., 2014). For organotypic cortex cultures, glial cells have been demonstrated to protect the neuronal tissue from mechanical damage (Schultz-Süchting and Wolburg, 1994; Schmidt-Kastner and Humpel, 2002). Also, organotypic cortex cultures show clear differences compared to *in vivo* cortex, such as an overall reduced connectivity due to a reduction in the 3$^{rd}$ dimension when preparing the brain slice taken into culture (Cäser and Schüz, 1992) or a change in glial protein expression (Staal et al., 2011). Organotypic cultures are typically prepared from newborn animals. Therefore, the cortical section of postnatal brain, which is taken for culturing is still immature, particularly with respect to the development of superficial layers. However, this immature cortex has benefited from structuring input during embryonic development, which has been shown to be important for somatotopic map formation, i.e. establishing a correspondence between body parts and brain regions (Antón-Bolaños et al., 2019). Therefore, the organotypic cortex culture should be best thought of as an *in vitro* system that has experienced a robust structural organization during embryogenesis and contains the blueprint for the organization of layered, cortical columns in isolation. The next section will summarize how structural self-organization continues as the cortex further matures in isolation and gives rise to several dynamical motifs of neuronal population activity.

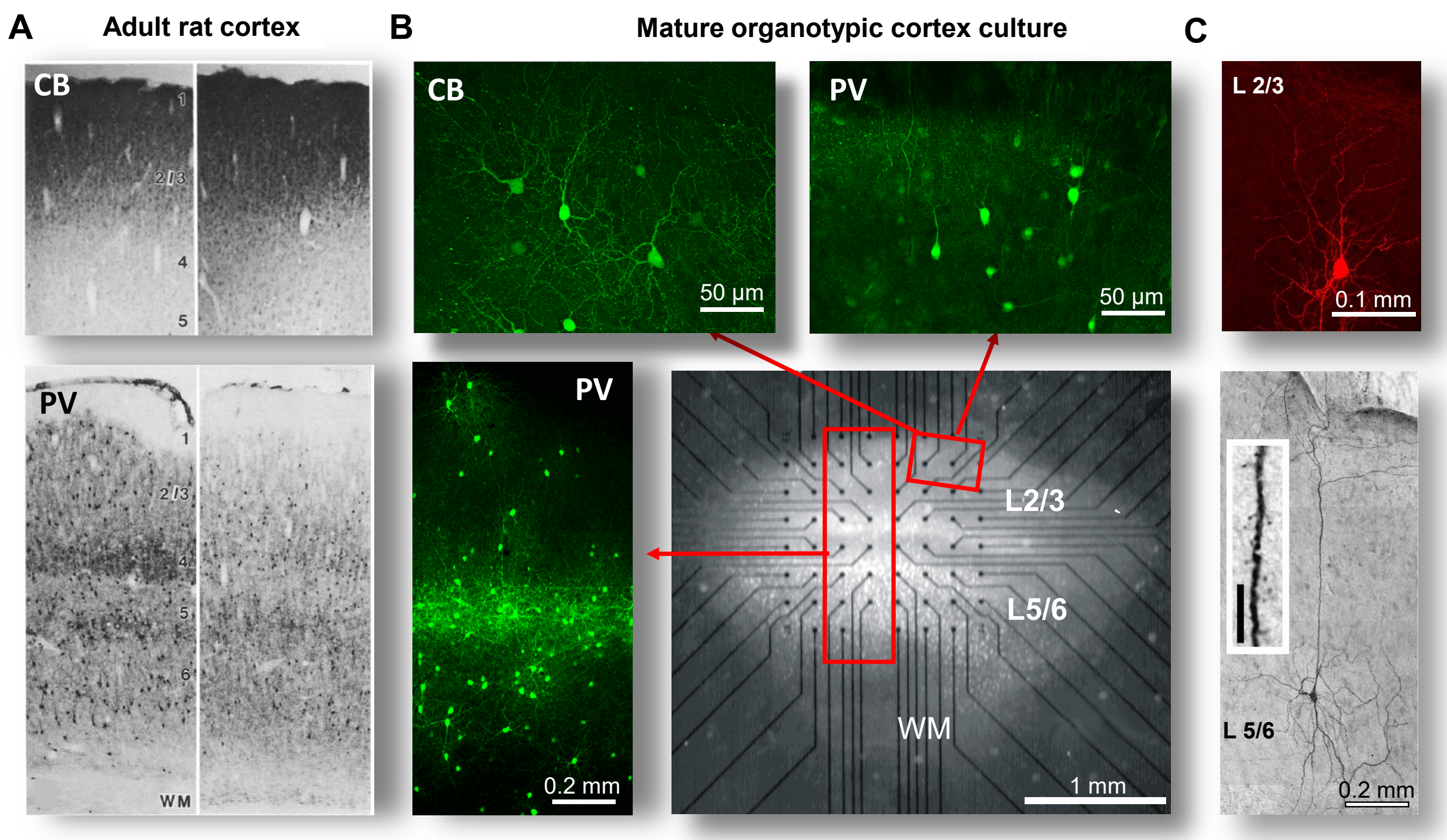

**Figure 1. Structural motifs of self-organization: cortical layers, pyramidal neurons, and interneurons in organotypic cortex cultures. A**, Coronal sections from the brain of adult rats showing somatosensory cortex (*left*) and motor cortex (*right*). Note high density of calbindin (*CB*) positive interneuron stain typical for superficial layers (*top*) and the layer dependent bands of parvalbumin (*PV*) positive interneurons in deep layers (*bottom*). **B**, Organotypic cortex culture after ~4 weeks grown on a planar multielectrode array (MEA). Note transparent healthy neural tissue covering ~2.6 mm$^2$ of the array at a thickness of ~100 – 200 µm and electrodes (*black dots*), conductors of the MEA. Composite images (*red rectangles*) indicating superficial layers (*L2/3*) that contain PV and CB positive interneurons and deep layers (*L5/6*) with their intense band of PV-positive interneurons. *WM*: white matter region. **C**, Typical cell body and dendritic morphology of pyramidal neurons from L2/3 and L5, the latter with their characteristically long and branched apical dendrite. *Inset*: Spiny dendrite typical for pyramidal neurons. [*A* reprinted with permission from (Van Brederode et al., 1991).] [*B* modified with permission from (Plenz and Aertsen, 1996a).] [*C* modified with permission from (Plenz and Kitai, 1998). Copyright 1998 Society for Neuroscience.]

### THE 1ST AND 2ND DYNAMICAL MOTIFS OF SELF-ORGANIZATION: UP-STATES AND NESTED θ/γ–OSCILLATIONS

The structural self-organization in organotypic cultures should parallel a self-organization of dynamical motifs found in the fully mature brain. One of these motifs, which is dominant in the electrocorticogram (ECoG) of humans in the awake state, is composed of transient, i.e. up to several seconds lasting, nested oscillations in the theta (θ: 8 – 12 Hz) and gamma (γ: > 25 Hz) range capturing the emergence of population synchrony at many local sites (Fig. 2A; (Canolty et al., 2006)). The nesting of high-frequency γ–oscillations to each θ–cycle has been proposed to be essential for working memory (Lisman and Idiart, 1995; Lisman and Jensen, 2013) and information transfer from lower to higher cortical areas (e.g., (André et al., 2015; Lundqvist et al., 2016)). In mature organotypic cortex cultures, detailed intracellular recordings demonstrated the presence of nested θ/γ–oscillations that arise during pronounced depolarizations that can last up to several second (Fig. 2B, C; (Plenz and Kitai, 1996; Klostermann and Wahle, 1999; Johnson and Buonomano, 2007)). This depolarization establishes the well-known dynamical motif of an 'up-state', which is typically defined as a prolonged period of self-sustained network excitation lasting from hundreds of milliseconds to several seconds.

The dominance of up-states supporting nested θ/γ–oscillations has several profound implications when studying SOC in isolated cortex preparations. First, it is well known that up-states, particularly prolonged ones (> 0.2 s), require stimulation of both the fast-acting (< 30 ms) AMPA-glutamate receptor and the slow-acting (> 50 ms) NMDA-glutamate receptor (Fig. 2D). The prolonged time course of the NMDA-glutamate receptor reduces the precision in action potential timing (Harsch and Robinson, 2000), suggesting that the scaffolding of precise spatiotemporal events requires alternative mechanisms, e.g., interneuron firing. Indeed, pyramidal neurons tend to fire sparsely during up-states, whereas interneurons fire reliably during almost every γ–cycle, a robust finding established in organotypic cortex cultures (Plenz and Kitai, 1996; Klostermann and Wahle, 1999; Czarnecki et al., 2012) and in acute cortex slices (Compte et al., 2008).

Second, the profound intracellular depolarization found in neurons during up-states indicates an overall increase in network activity. However, the up-state depolarization should not be equated with a higher excitability of individual neurons, which is implicitly assumed in neuronal models that do not take intracellular membrane conductance changes into account, i.e., due to synaptic inputs (Bernander et al., 1991). On the contrary, individual neurons significantly change how they respond to additional input during up-states (Petersen et al., 2003; Czarnecki et al., 2012; Reig et al., 2015); this change is effected by a rather expansive combination of a decrease in neuronal input resistance (Monier et al., 2008), a shortened synaptic integration window (Bernander et al., 1991), transient changes in the balance of excitatory to inhibitory (E/I) synaptic transmission (Haider et al., 2006), active dendritic conductances (Ness et al., 2016), a critical slowing down of the threshold to action potential generation (Meisel et al., 2015) and other mechanisms (for further reading see (Fregnac et al., 2004)). Few neuronal simulations take these changes during the build-up of network activity into account (Markram et al., 2015; Dura-Bernal et al., 2019), potentially limiting insights that can be gained into these dynamical motifs from less biophysically oriented modelling.

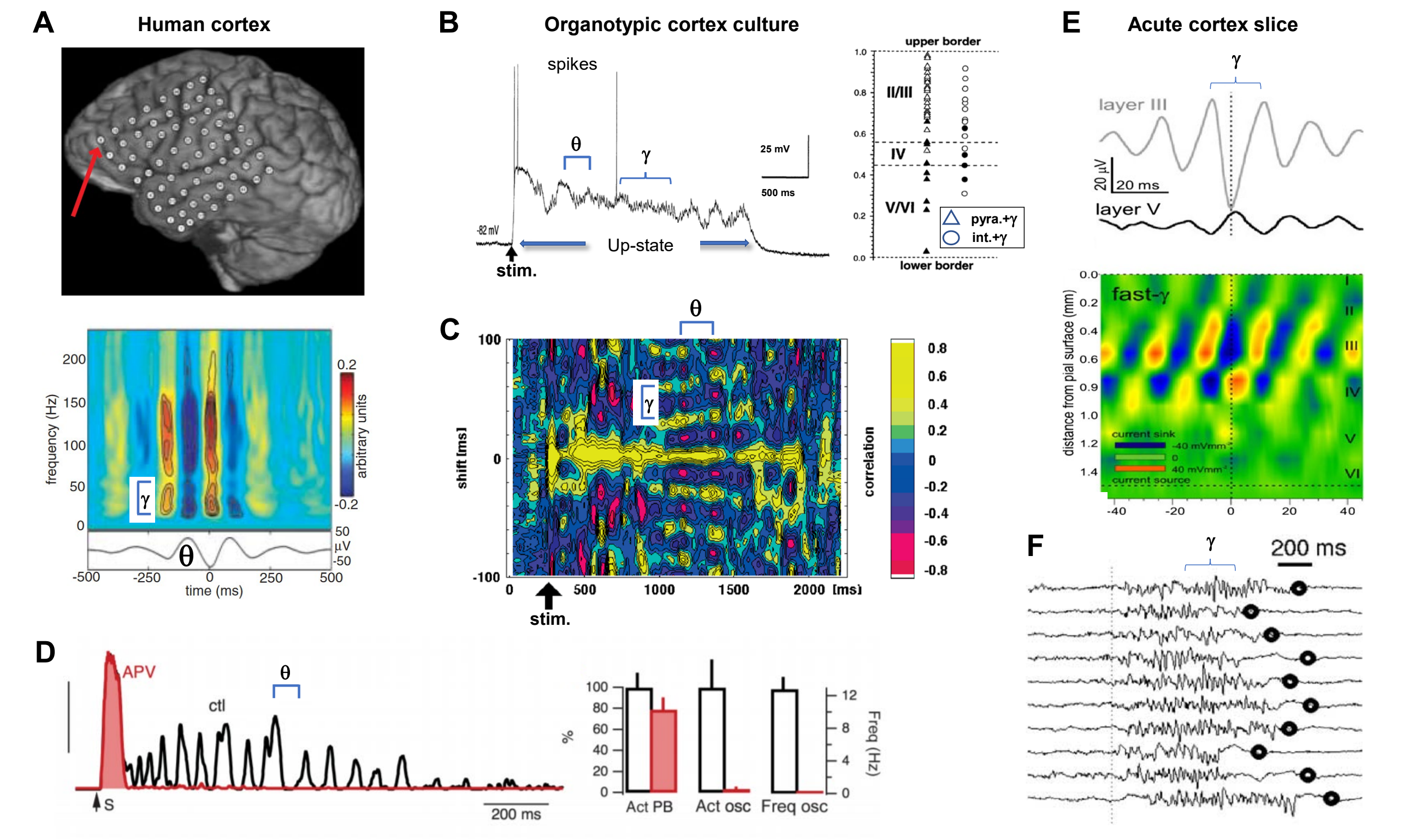

**Figure 2. The 1st & 2nd dynamical motif of dynamical self-organization: up–state and nested θ/γ–oscillations in organotypic cortex cultures and acute cortex slices. A**, The human brain displays distinct periods of nested θ/γ–oscillations at rest and during behavior. *Top*: Electrode array on the cortex surface (*circles*) recording the electrocorticogram. *Bottom*: Corresponding power spectrum of fast γ–oscillations (> 25 Hz; *top*) phase-locked over several cycles of a θ–oscillation (*bottom*). [Reprinted with permission from (Canolty et al., 2006).] **B**, In mature organotypic cortex cultures, neuronal activity self-organizes into up-states with nested θ/γ–oscillations in superficial layers. *Left*: Time course of the intracellular membrane potential for a pyramidal neuron in response to millisecond-lasting, electrical shock stimulation (*stim; arrow*). Note self-sustained up-state with nested θ/γ–oscillations and sparse occurrence of action potentials (*spikes*). *Right*: γ–oscillations are found in pyramidal neurons (*triangles*) and interneurons (*circles*) of superficial (*open*), but not deep layers (*filled*). **C**, Cross-correlation over time in the membrane potential of two pyramidal neurons in response to electrical shock stimulation (*stim; arrow*). Note maintained phase-locking of nested θ/γ–oscillations during the up-state with drop in γ–frequency creating a fan-out pattern [Subpanels *B*, *C* modified from (Plenz and Kitai, 1996).] **D**, Self-sustained up-states with θ–oscillations require recurrent, excitatory network connections. Blocking the excitatory NMDA glutamate-receptor with the antagonist APV only leaves an initial, short-lasting direct response (organotypic cortex culture). (*red*). *Left*: Population activity time course to electric shock stimulation (arrow; *S*). *ctl*: control. *Right*: APV only slightly reduces the number of up-states (*left most bars*) but blocks the emergence of oscillations at θ–frequency (*middle and right most bars*). *Act*: control. [Reprinted with permission from (Czarnecki et al., 2012).] **E**, In the acute cortex slice, γ–oscillations emerge in superficial layers (CSD analysis). [Reprinted with permission from (Oke et al., 2010).] **F**, Acute cortex slice from adult ferret with spontaneous, self-sustained period of fast γ–oscillations in the LFP. [Reprinted with permission from (Compte et al., 2008). Copyright 2008 Society for Neuroscience.] *Oscillation identifiers* have been added to some subpanels for clarity.

Third, nested θ/γ–oscillations during the up-state are not blocked by the gap junction blocker carbenoxolone (Gireesh and Plenz, 2008) and the activity propagates relatively fast, with a velocity > 50 mm/s (Beggs and Plenz, 2003). These findings support the view that nested θ/γ–oscillations originate in superficial layers from synaptic interactions between local interneurons and pyramidal neurons (Plenz and Kitai, 1996). These nested oscillations are therefore considered to differ from so-called slow oscillations, which *in vivo* can be induced by deep but not superficial layer stimulation (Beltramo et al., 2013). Slow-oscillations were shown *in vitro* to originate in deep layers and to propagate significantly slower than θ/γ–oscillations locally, yet they were shown to contribute to up-state initiation in superficial layers (Sanchez-Vives and McCormick, 2000; Wester and Contreras, 2012; Capone et al., 2017).

The propensity of isolated cortex to produce up-states and nested oscillations is demonstrated in the acute cortex slice as well, in which tissue is studied within hours after being taken from the adult brain. In the acute slice, synchronized nesting during up-states can be induced by an external, pharmacological stimulation which includes direct neuronal depolarization through excitatory glutamate receptors in combination with the neuromodulator acetylcholine (Fig. 2E, F; (Buhl et al., 1998; Compte et al., 2008; Yamawaki et al., 2008)). Current-source density (CSD) analysis, which tracks the spatial location of neuronal current generation (Nicholson and Freeman, 1975; Mitzdorf, 1985; Plenz and Aertsen, 1993), demonstrates that nested θ/γ–oscillations originate in superficial layers 2/3 in both the acute cortex slice (Fig. 2E; (Oke et al., 2010)) and organotypic cortex culture (Fig. 3). Developmentally, these dynamical motifs occur in organotypic cultures with a similar time course compared to *in vivo*, specifically when co-culturing cortex with midbrain regions, which provide the neuromodulator dopamine (Fig. 3; (Gireesh and Plenz, 2008)). In summary, isolated cortex preparations demonstrate the autonomous emergence of two dynamical motifs in superficial layers of cortex: up-states and nested oscillations.

### THE 3RD DYNAMICAL MOTIF OF SELF-ORGANIZATION: NEURONAL AVALANCHES

Until now, the two dynamical motifs of up-states and nested oscillations have been treated from the point of view of averages. In this view, an up-state is approximated as a binary transition between two network states that differ in overall activity and oscillations are treated to be spatiotemporally stationary. These views are ill-equipped to capture spatiotemporal propagation in synchronized cortical activity as well as the spatiotemporal variability encountered in spontaneous or evoked instantiation of synchronous activity.

In contrast, the 3rd dynamical motif of self-organization, neuronal avalanches (Beggs and Plenz, 2003) emphasizes both spatiotemporal propagation, as well as variability in cortical synchronization. In that respect, avalanches are related to the spatially compact, wave-like propagation of cortical activity (Ermentrout and Kleinfeld, 2001; Rubino et al., 2006; Takahashi et al., 2011) as well as the concept of 'synfire chains', spatiotemporally selective cascades of neuronal firing proposed by Abeles (Abeles, 1991). Neuronal avalanche dynamics introduces several major concepts with respect to propagation and variability in cortical synchronization. First, avalanche dynamics quantifies synchronization within a period of duration $\Delta t$ and successive occurrences of synchronization in near future time periods. It therefore covers ‘instantaneous’ as well as propagated synchrony (see Fig. 4). Second, avalanche dynamics exhibits scale-invariance in space and time, which introduces power laws as the statistical measure of variability and the concept of critical branching (see Figs. 4, 5). Third, avalanche dynamics allows for the decomposition of propagated synchrony into ‘coherence potentials’, a previously unknown concept in cortical dynamics for information transfer (Fig. 6). Fourth, avalanche dynamics lifts the idea of *one* particular spatiotemporal pattern to that of ‘avalanches of avalanches’, which serves as a set of very specific predictions of how spatiotemporal synchronization events in the cortex are linked to each other in sizes and time (see Fig. 10). Finally, avalanche dynamics introduces quantitative and absolute measures to study optimization in cortical networks (see Figs. 11, 12). We will elaborate on these major

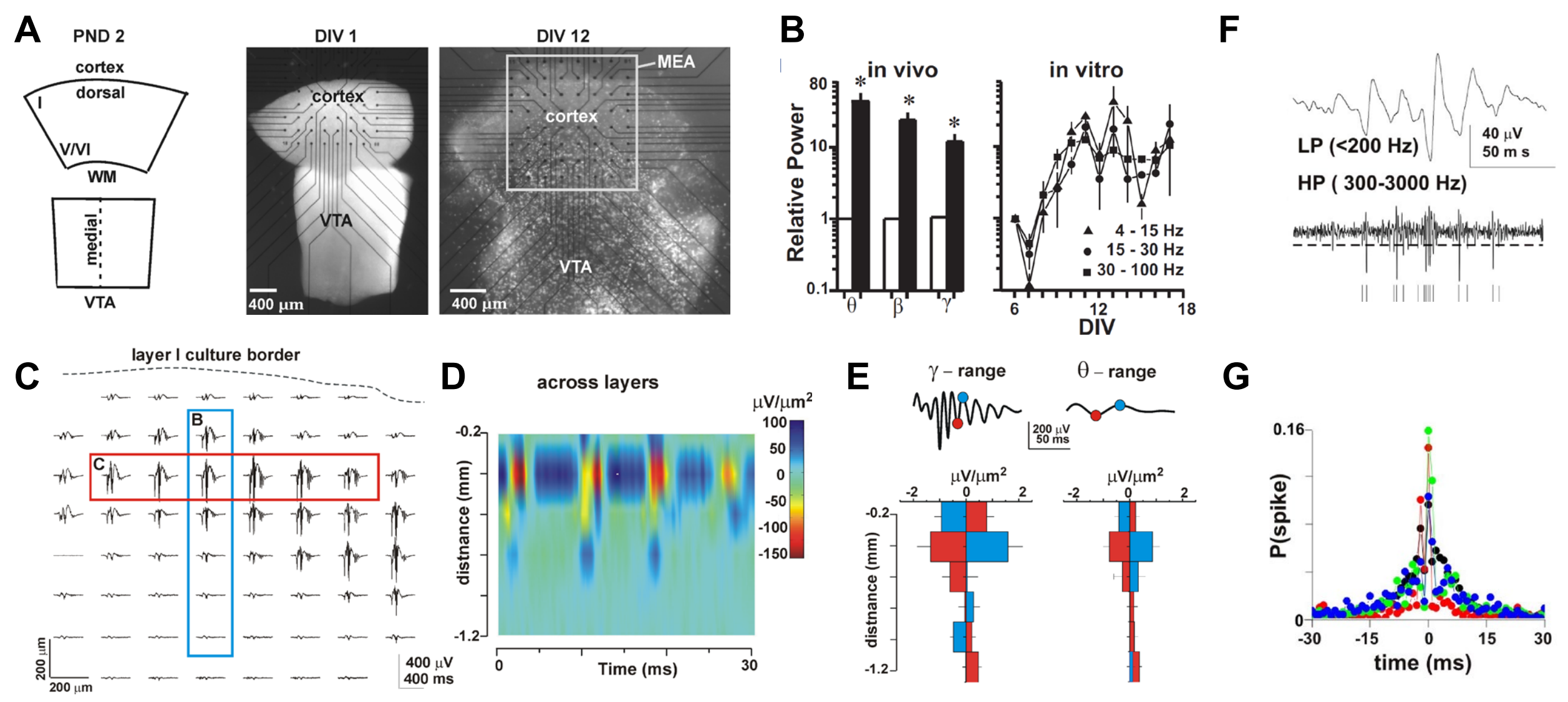

**Figure 3. Self-organization into nested θ/γ–oscillations in organotypic cortex cultures during the 2nd postnatal week when superficial layers mature**. **A**, Coronal sections of cortex and ventral tegmental area (*VTA*) are combined and grown on the MEA over two weeks in culture. *Left*: Sketch of coronal cortex slice and midline crossing midbrain region containing the VTA at postnatal day (*PND*) 1 – 2, when taken into culture after 1 day *in vitro* (*DIV*; *middle*), and about 2 weeks later (*DIV* 12; *right*). Note flattening of the culture visible by the increased transparency and expansion of the dorsal tissue on the MEA as superficial layers develop (light microscopic image). **B**, Nested oscillations increase in power from 1st (*white*) to 2nd (*black*) postnatal week *in vivo* (*left*), a developmental time course mirrored in organotypic cortex-VTA co-cultures (*right*). **C**, Spontaneous nested θ/γ–oscillations distribute along the dorsal part of the cortex within superficial layers (*red rectangle*; single culture). **D**, Corresponding CSD (across *blue rectangle* in C) demonstrates θ/γ–oscillations to originate from synaptic sources (*sinks*) within superficial layers. CSD density with distance from upper culture border over time. **E**, Summary of average CSD with distance from dorsal border for 7 cultures separated into θ– and γ–oscillations activity. **F**, Separating the broadband LFP into spike-information carrying high-frequency band (*HP*) and the population activity containing low-frequency band (*LP*) demonstrates local spiking (vertical bars) to be phase-locked to γ–oscillations. **G**, Spike probability is highest at the negative peak of the γ–cycle (nLFP) at t = 0 (color code shows 4 cultures). [Reprinted with permission from (Gireesh and Plenz, 2008). Copyright 2008 National Academy of Sciences.]

conceptual changes in studying cortical synchronization in the following sections.

We start with the basic definition of avalanches using the comparative *in vivo* and *in vitro* study on the developmental emergence of neuronal avalanches in superficial layers of cortex (Figs. 3, 4). Gireesh and Plenz (2008) used multielectrode array (MEA) recordings to demonstrate the embedding of avalanches into ongoing nested oscillations. Using a simple threshold crossing approach, they extracted the time and amplitude of negative peak deflections in the LFP (nLFP) at each electrode in order to identify location, time, and the size of short-lasting, synchronized activity in a local group of neurons (Fig. 3F, G; Fig. 4A; Plenz 2012; Petermann et al 2008). Contiguous time periods with nLFPs were defined as avalanches (Fig. 4A, bottom), which resulted in a large variety of different patterns. The size of these patterns, here defined as the absolute sum of nLFPs distributed according to a power law up to a cut-off, serve as the hallmark of neuronal avalanches (Fig. 4C). This power law was also found when defining avalanche size by the number of threshold crossing electrodes (Gireesh and Plenz, 2008), which approximates the spatial extent of avalanches (Yu et al., 2014). The embedding of avalanches in nested oscillations clearly emerges in cortex-VTA co-cultures, with avalanche size distributions exhibiting a precise power law up to the cut-off (Fig. 4C, D). These findings established that the complex developmental signature of avalanches and nested oscillations *in vivo* develops autonomously in organotypic cortex cultures with a similar developmental time course, i.e. it is established towards the end of the 2nd week postnatal, in the absence of any structuring sensory input or motor output (Fig. 4C, D). The precise match of the power law in avalanche sizes with slope of -3/2 that emerges from the variability of nested θ/γ–oscillations is not a statistical coincidence. Besides both dynamical motifs being highly sensitive to fast inhibition via the $GABA_A$ receptor and slow excitation via the NMDA–glutamate receptor, this co-existence required fine-tuning via the dopamine $D_1$–receptor. Specifically, when the dopamine $D_1$– but not $D_2$–receptor was blocked, nested oscillations continued to emerge, yet the resulting nLFP cascades now exhibited a much steeper size distribution (Gireesh and Plenz, 2008). This regulation of avalanche size distributions to a slope of -3/2 as a function of NMDA/$D_1$–receptor co-stimulation has been confirmed for superficial layers in acute slices of prefrontal cortex taken from two months old, adult rats (Stewart and Plenz, 2006; Bellay et al., 2021) (*cf.* Fig. 11). Recent analysis *in vivo* in the prefrontal cortex of awake nonhuman primates further confirmed this precise relationship between avalanche dynamics and γ–oscillations (Miller et al., 2019).

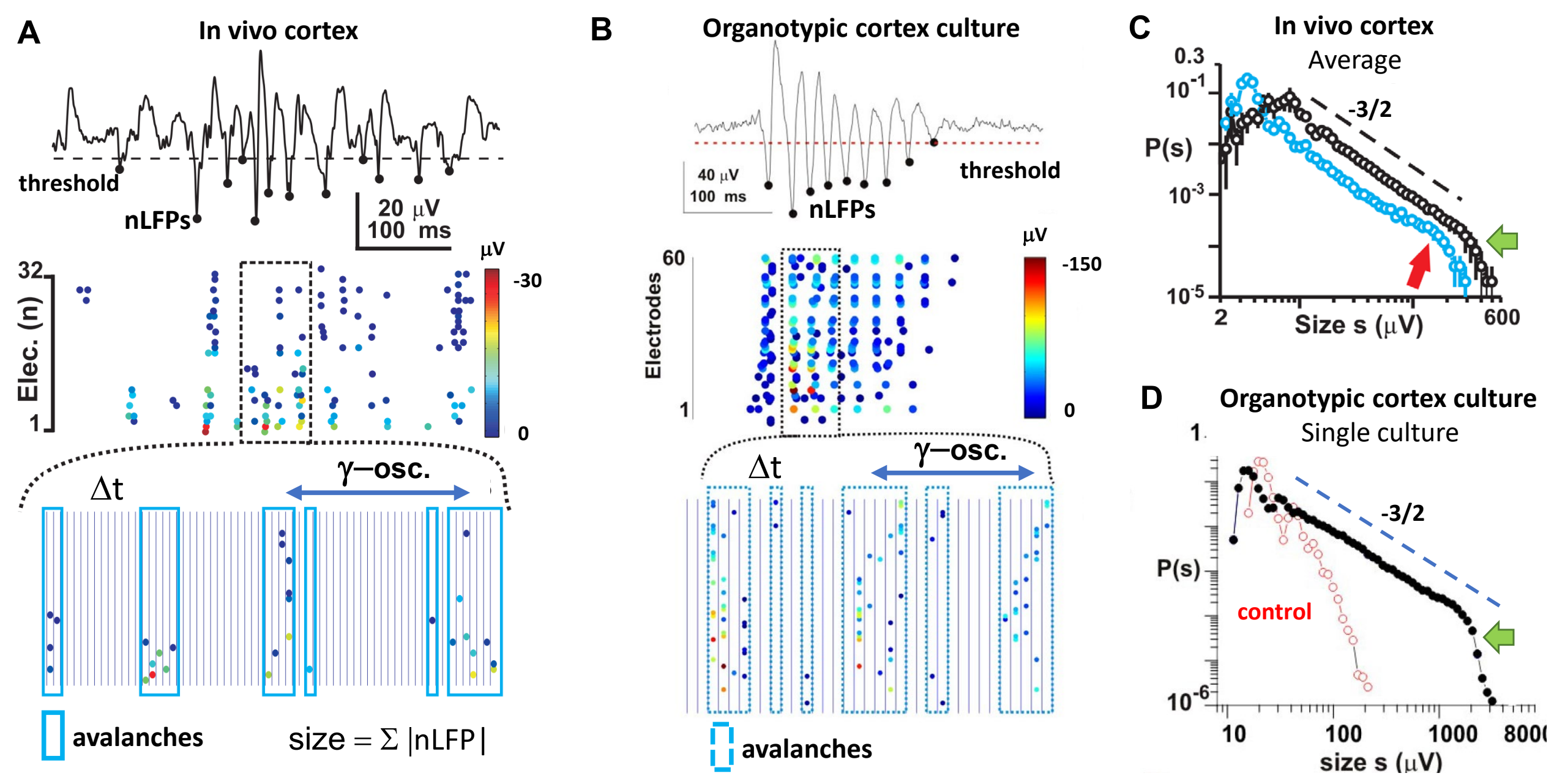

**Figure 4. The 3rd dynamical motif of self-organization: neuronal avalanches embedded in nested oscillations. A**, Nested θ/γ–oscillations in the superficial layer of a 2-week old rat *in vivo* organize as neuronal avalanches that predominantly emerge at oscillation cycles. *Top*: Single cortical electrode with continuous LFP from which negative peak deflections (nLFP; *filled circles*) are identified after threshold crossing. *Middle*: Corresponding raster of nLFPs on the full 32-MEA exhibiting columnar, variable bouts of synchronization in space and time. nLFP peak amplitudes are color coded. *Bottom*: Expanded view of identified avalanches (*blue rectangles*) during each γ–oscillation period plotted in discrete time steps of duration *Δt* (*vertical lines*). Size of an avalanche is defined as the sum of absolute nLFP amplitudes within each avalanche. **B**, As in *A* but for a transient γ–oscillation in an organotypic culture. **C**, At the end of the 2nd postnatal week (*black*), avalanches distribute in size from 10 – 300 µV followed by a cut-off (*green arrow;* n = 7 animals). *Broken line*: slope of -3/2 as guide to the eye. Size distribution is slightly bimodal (*red arrow*) at the beginning of the 2nd postnatal week (*blue*; n = 5 animals). **D**, Power law in avalanche size distribution for a single culture and exponential size distribution obtained from time-shuffled controls in which spatial and temporal correlations among nLFPs are destroyed. [Reprinted with permission from (Gireesh and Plenz, 2008). Copyright 2008 National Academy of Sciences.]

We note that the definition of neuronal avalanches, originally introduced by Beggs and Plenz (2003) using the LFP, requires that each local site exceeds a minimal activity threshold. Using a neuronal network model, Poil et al. (2012) adopted a scheme in which the summed spiking activity within *Δt* of the finite-size network is required to exceed a population threshold. This latter definition is very similar to a threshold applied to the LFP, as will be argued in more detail below (*cf.* Fig. 13). It potentially introduces linear terms in certain scaling relationships as pointed out by Villegas et al. (2019). As for statistical tests demonstrating the presence of a power law in avalanche size distributions, we refer to Yu et al. (2014) for a more detailed discussion.

To summarize, *in vivo* experiments in rodents and nonhuman primates, as well as developmentally well controlled *in vitro* experiments using organotypic cortex cultures and acute cortex slices, demonstrate a precise regulation between up-states, nested oscillations and neuronal avalanches that involves fast GABA-mediated inhibition, slow-glutamate mediated excitation and the neuro-modulator dopamine.

## TEMPORAL AND SPATIAL SCALING LINKS SIZE DISTRIBUTION SLOPE -3/2 TO A CRITICAL BRANCHING PARAMETER FOR NEURONAL AVALANCHES

The identification of avalanches and their implication for SOC has been a particular challenge from an experimental point of view. Besides the structural constraints of superficial layers and developmental period that must be considered, there are additional aspects specific to the emergence of neuronal avalanches themselves that are of importance. These points will be addressed in the following. The original identification of neuronal avalanches (Beggs and Plenz, 2003) involved numerous scaling controls to demonstrate that power laws identified in propagated neuronal activity were robust to obvious choices in the experimental setup. Specifically, tracking the spatiotemporal spreading of an avalanche using discrete, spatial sensors such as MEAs, requires the appropriate choice of a discrete time interval *Δt* (Fig. 5A). This choice of *Δt* is imposed by the average, finite propagation velocity $<v>$ for neuronal activity in the system and the introduction of a discrete sampling distance of *Δd* by the MEA. Three observations laid the groundwork that established the power law in avalanche sizes with a slope $\alpha = -3/2$. First, increasing $\Delta t$, while keeping $\Delta d$ constant, led to a shallower slope $\alpha$ without change in power law shape. This dependency of $\alpha(\Delta t)$ itself is approximated by a power law, allowing for scaling collapse *in vitro* (Fig. 5B, C) and *in vivo* (Petermann et al., 2009). Second, when changing $\Delta d$ and accordingly adjusting $\Delta t = \langle v \rangle * \Delta d$, a robust size exponent of $\alpha = -3/2$ was obtained (Fig. 5C). Third, the cut-off of the power law was simply a function of the finite MEA size and did not change $\alpha$ itself (Fig. 5D). Importantly, when cultures were made more excitable, by reducing inhibition in the system using pharmacological means, the avalanche size distribution changed from a power law to a bimodal distribution exhibiting an initial steep slope close to -2 and pronounced system size peak indicative of all-or-none, system-wide population events (Fig. 5E). This latter separation into local, non-propagated events and large system-wide synchronization, exhibits the phenotype of a 1$^{st}$ order discontinuous phase-transition. We will point out in detail in subsequent sections that these scaling operations are not robustly observed in dissociated culture experiments, where an increase in $\Delta t$ typically *steepens* the initial slope and uncovers a bimodal cascade size distribution (see below for details; *cf.* Fig.13), which is more in line with a hyperexcitable system.

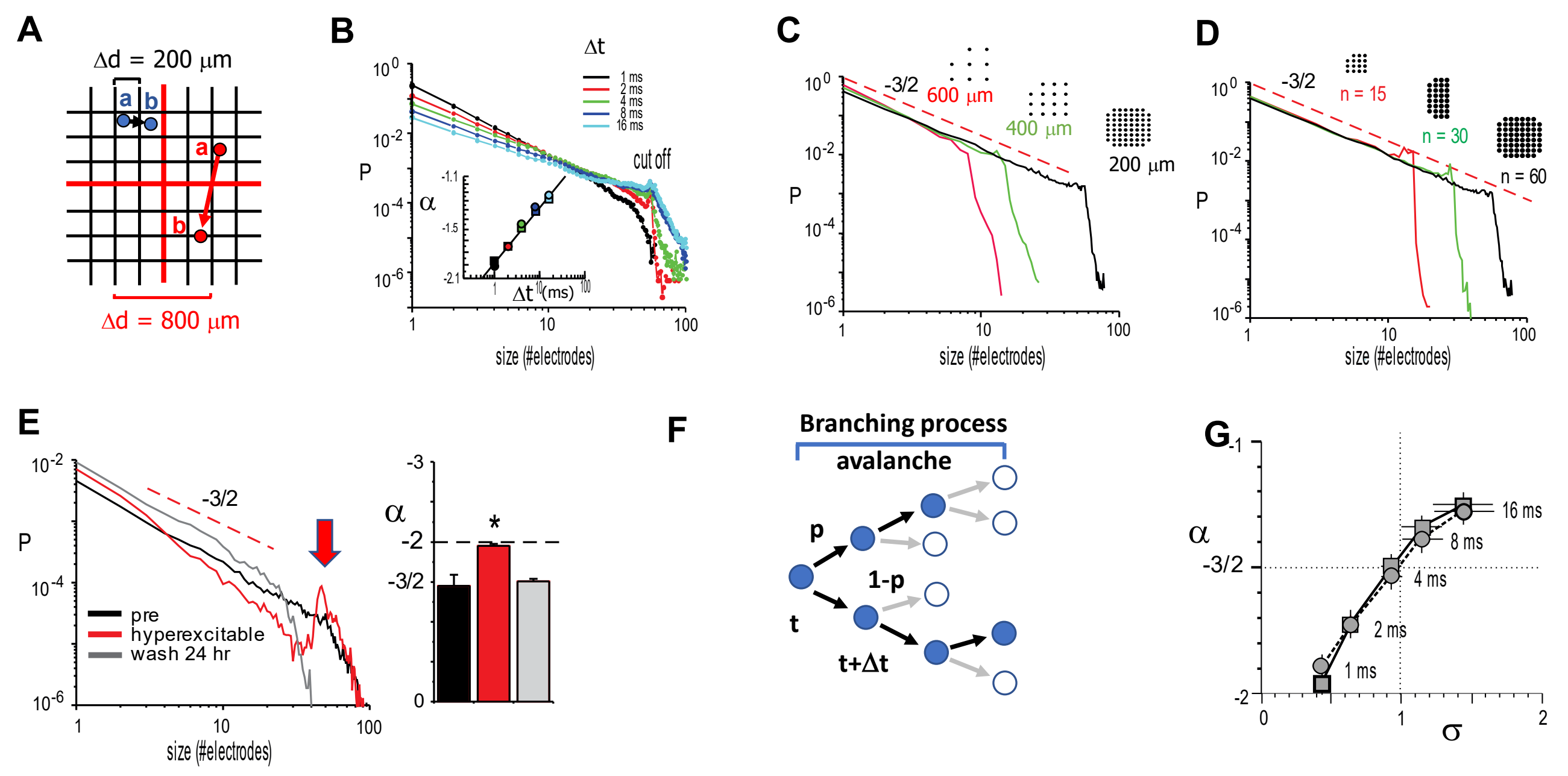

**Figure 5. Temporal and spatial scaling links size distribution slope -3/2 to a critical branching parameter for neuronal avalanches.** **A**, When identifying spatial propagation of activity over time (a → b; blue & red circles), the distances between neighboring electrodes on a multielectrode array introduce a discrete spatial scale $\Delta d$. For different spatial scales (grids: *black*: 200 µm; *red*: 800 µm) and a finite propagation velocity $\langle v \rangle$, the time $\Delta t$ to wait in order to identify propagation towards an electrode is approximately $\Delta t \cong \langle v \rangle * \Delta d$. **B**, In organotypic cortex cultures, at fixed $\Delta d = 200$ µm, an *increase* in $\Delta t$ results in a power law of avalanche size with a *shallower* exponent $\alpha$ due to preferential concatenation of avalanches (avalanche size based on number of active electrodes). Note the absence of any change in the power law form itself. *Inset*: Average change in $\alpha$ with $\Delta t$ over all cultures and size based on electrodes (*circles*) or nLFP amplitudes (*squares*). **C**, A change in $\Delta d$ by accordingly omitting electrode rows and columns (*insets*), maintains the size distribution slope of -3/2 (*broken line*) if correspondingly increasing $\Delta t$ to $\Delta t \cong \langle v \rangle * \Delta d$. **D**, Finite-size scaling using compact sub-arrays (*insets*) only affects the power law cut-off, but not power law slope $\alpha$. *Broken line*: slope of -3/2. **E**, Reducing inhibition pharmacologically using picrotoxin leads to hyperexcitable neuronal cultures (*red*), resulting in a 'supercritical' phenotype with an initial, steep slope close to ~-2 and a preference for large, i.e. system-wide propagated population events (*red arrow*). *Right*: Quantification of change in $\alpha$ from -3/2 → -2 when reducing inhibition (*red bar*). **F**, The unfolding of an avalanche in a network viewed as a branching process. In this sketch, activity at an initial site at time $t$ can induce activity within $t+\Delta t$ at another site with probability $p$ or fails to induce activity at a new site with probability $1 - p$. Each site exhibits a potential branching to two new sites. **G**, Experimental demonstration that avalanche dynamics crosses the critical point $(\sigma, \alpha) = (1, -3/2)$ predicted for a critical branching process with change in $\Delta t$ at fixed $\Delta d$. Avalanche size definition (*squares, circles*) as in *B*, inset. [Subpanels *B – E*, *G* reprinted with permission from (Beggs and Plenz, 2003). Copyright 2003 Society for Neuroscience.]

The original work by Beggs and Plenz (2003) provided the first insights of a critical branching process as a proxy to understand avalanche dynamics in cortical networks as well. A memoryless branching process captures the probability of an initial event to spawn future events at new sites (Harris, 1963). The corresponding branching parameter, $\sigma$, quantifies the average ratio of next generation to currently active sites (Fig. 5F). For random neighbors and $\sigma$ = 1, the resulting size distribution from such an unbiased or critical branching process exhibits a power law with slope of -3/2 and can be analytically linked to the self-organized critical sandpile (Christensen and Olami, 1993). In line with these basic expectations, it was found that $\sigma$ is close to 1 and $\alpha$ = -3/2 for neuronal avalanches at $\Delta t = \langle v \rangle * \Delta d$ (Fig. 5G). These findings introduced branching processes as a promising entry point to study avalanche generation.

These original scaling operations for avalanches involved > 10 hr of continuous recordings *in vitro*, which is difficult to achieve under standard experimental conditions. Recently, Miller et al. (Miller et al., 2019; 2021) extended this scaling analysis of LFP based avalanches. They identified a scaling exponent of 2 for avalanche waveform and mean size vs. duration relationship in line with predictions for a critical branching process. We also note that LFP avalanches show nearest-neighbor propagation and typically involve no loops (Yu et al., 2014). The precise identification of scaling exponents for neuronal avalanches and the conditions under which they are robust is currently an intense field of research. Several alternative processes, both critical and non-critical, have been suggested to produce size exponents close to -3/2 (for further reading see e.g. (Martinello et al., 2017; Touboul and Destexhe, 2017)). In the following sections, we will focus on additional dimensions of neuronal avalanche dynamics that go beyond these basic scaling relationships. Importantly, the presence of a power law in avalanche sizes and a critical branching parameter of $\sigma$ = 1 is linked to several distinct aspects in the emergence and propagation of neuronal activity.

## THE 4$^{TH}$ DYNAMICAL MOTIF OF SELF-ORGANIZATION: THE COHERENCE POTENTIAL

In the previous section, the scaling relationship between the temporal and spatial resolution was reviewed. The third free parameter in assessing avalanche dynamics is the threshold, $\lambda$, at which a local site is considered to carry significant activity. For LFP based avalanches, this threshold is typically chosen to be around 3 SD of the fluctuations in activity at each site and it has been shown in numerous studies that the presence of a power law is rather robust to the threshold chosen, assuming it is reasonably outside of baseline noise (Beggs and Plenz, 2003; Petermann et al., 2009; Tagliazucchi et al., 2012). Yet, when systematic evaluations of threshold effects were conducted within a regime of robust power law scaling, it was found that avalanche dynamics implicitly contains a local synchrony threshold that identifies a subclass of avalanches *in vivo* as well as *in vitro*: the coherence potential (Fig. 6; (Thiagarajan et al., 2010; Plenz, 2012)).

Coherence potentials constitute avalanches with nLFPs above a minimal amplitude threshold typical about ~3 SD of the ongoing LFP fluctuations (Yu et al., 2011). Both avalanches and coherence potentials form power laws in size distributions that are indistinguishable by simple thresholding (Fig. 6A). Only when the waveform of nLFPs is explicitly taken into account, is a sigmoidal function identified separating the high-fidelity activity propagation regime that constitutes the relatively small number of coherence potentials from that of all other avalanches (Fig. 6B – D).

The identity of nLFP waveform correlates with the identity of local spike sequences across different cortical locations (Thiagarajan et al., 2010), suggesting that coherence potentials confer the exact temporal activity of local neuronal firing over wide distances of cortex. In the human ECoG, coherence potentials were found to initiate finger tapping (Parameshwaran et al., 2012). The emergence of coherence potentials in cortical networks with avalanche dynamics has been compared to the emergence of 'gliders' in cellular automata and hypothesized to be a vehicle of information

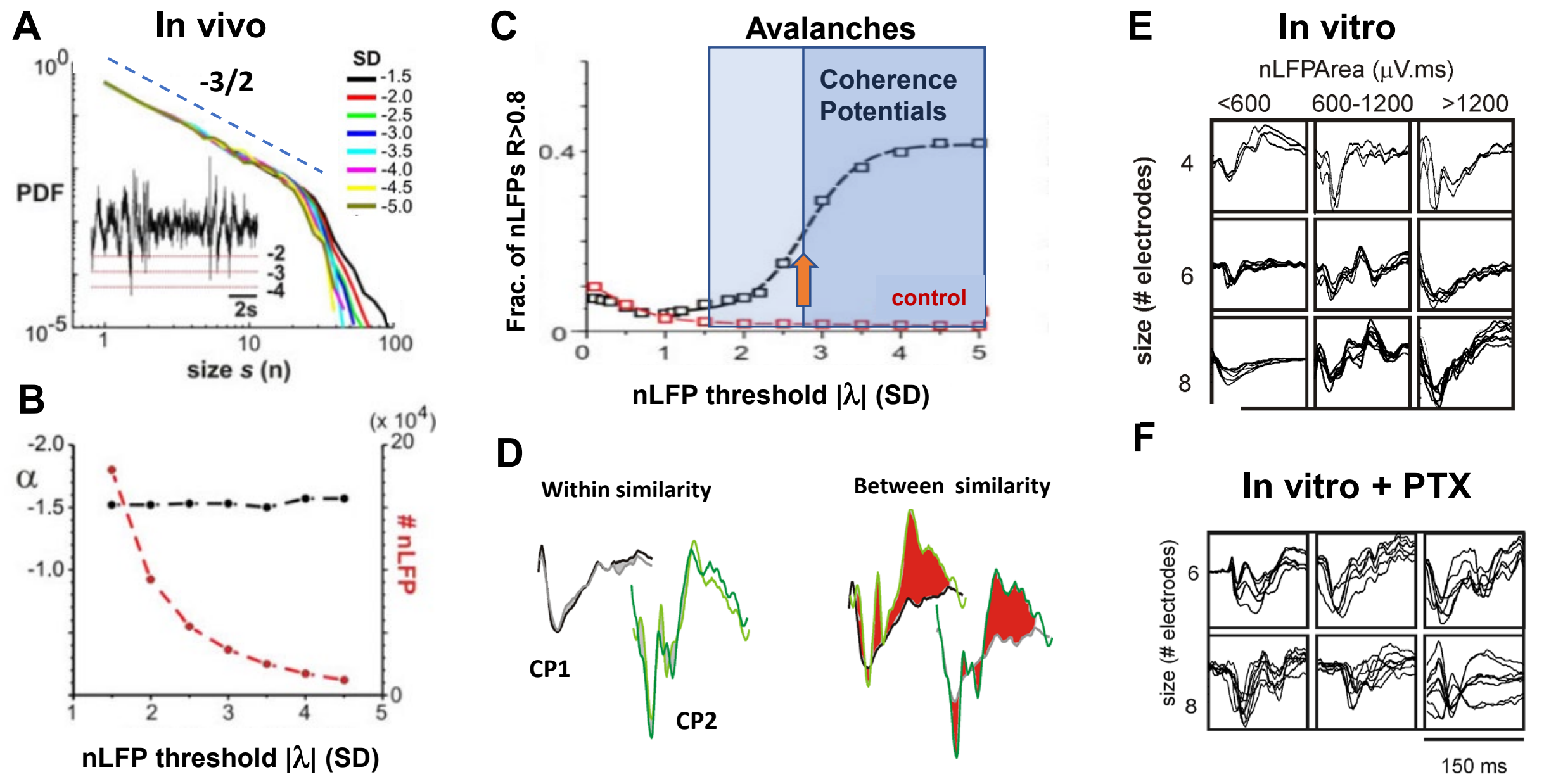

**Figure 6. The 4th dynamical motif of self-organization: the coherence potential. A**, LFP avalanches from ongoing activity in nonhuman primate show a robust power law in sizes for a range in threshold $\lambda$ at which nLFPs are detected. *Inset*: example LFP trace and nLFP detection for 3 different values of $\lambda$. *SD*: Standard deviation. *Broken line*: slope of -3/2. **B**, The number of nLFPs > $\lambda$ drops several orders of magnitude with increase in $\lambda$ while the power law in sizes remains robust with slope α close to -3/2. **C**, The similarity in nLFP waveforms within avalanches rapidly increases for $\lambda$ > ~2.7 SD (*orange arrow*). This threshold-dependent sigmoidal increase in within-nLFP similarity identifies a relatively small number (~20%) of nLFPs that constitute coherence potentials (cf. *B*). *Control*: Expected change in similarity after random temporal shifts in the LFP. **D**, Complex nLFP waveforms within a coherence potential are highly similar. Example comparison for two subsequent coherence potentials at two sites. Within similarity (*gray/black*; *light/dark green*) is high for each coherence potential (*CP*), whereas between similarity is very low with large waveform deviations highlighted in *red* for visual contrast. **E**, Coherence potentials spontaneously form *in vitro* in organotypic cultures. Overplot matrix of waveform examples for coherence potentials of size 4 – 8 for nLFP areas in three size categories. **F**, Under disinhibited conditions (*PTX*), coherence potentials are absent. [Reproduced with permission from (Plenz,

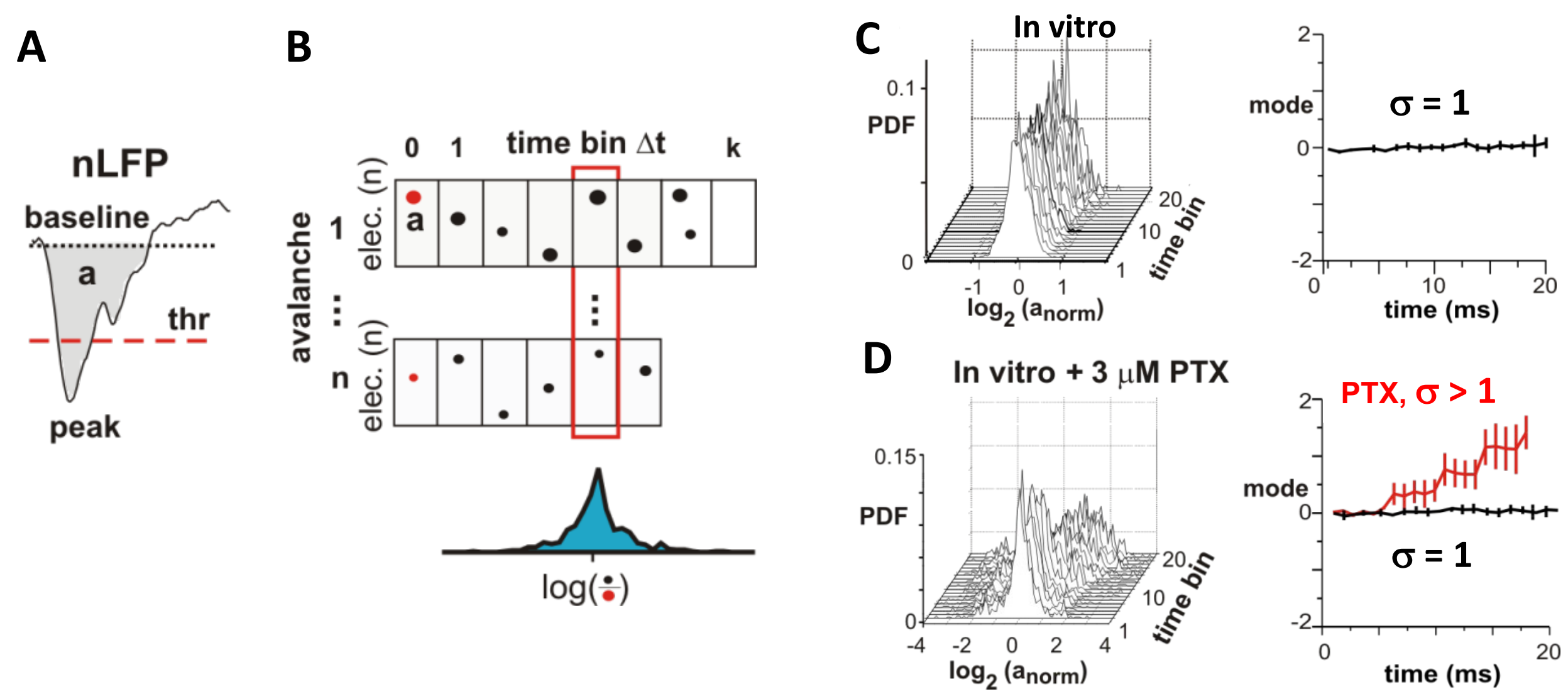

**Figure 7. Critical branching parameter estimation for coherence potentials. A**, Area *a* and waveform of a local nLFP identified by threshold (*thr*) crossing. **B**, nLFP areas normalized by initial nLFP area (*red dots*) distribute around a median value (*blue distribution*) after *n* time bins of duration *Δt*. **C**, *In vitro* avalanches reveal distribution around log(1) = 0 mode demonstrating that, as an avalanche unfolds, nLFPs on average neither grow nor decay in area in line with expectations for a critical branching process with $\sigma = 1$. **D**, In organotypic cortex cultures that are made more excitable by slightly reducing inhibition (*PTX*), an expansion of nLFP area with time from cascade initiation is found in line with prediction for a supercritical branching process with $\sigma > 1$. These experiments demonstrate that a transition from a power law in sizes to a bimodal size distribution under reduced inhibition changes the system from critical to supercritical dynamics. [Reproduced with permission from (Plenz, 2012).]

transfer within cortex at the network level (Plenz, 2012).

The waveform identity in coherence potentials could be expanded to area identity *in vitro* and to *in vivo*. By grouping nLFPs in coherence potentials into different size categories, waveform similarity within these categories was established and shown to break down when the network was disinhibited (Fig. 6E, F). This demonstrates that coherence potentials are actively regulated by the network through the E/I-balance. Coherence potentials were shown to demonstrate initial group size conservation as well. Specifically, it was demonstrated that the area of nLFPs, which participate in a single coherence potential, does not grow nor decay on average as the coherence potential unfolds, a finding that is independent of the size of the initiating nLFP (Fig. 7A – C). This property of preserving the local group size initiating an avalanche was lost when the cortex was even mildly disinhibited, upon which propagated activity displayed a within-cascade explosive growth (Fig. 7D). This particular approach extends the original identification of the critical branching parameter (Beggs and Plenz, 2003), which was estimated by the ratio between the number of nLFPs in the second ('descendants') and first ('ancestors') time bin of an avalanche. The analysis in Fig. 7 is more complete by including nLFP area and waveform and considers all avalanches in their full duration. The critical branching parameter is reflected in the finding that normalized distributions have a stable mode of 1, i.e., log(1) = 0, for up to 20 ms of propagation, which typically covers the full area of recording. The area of the nLFP correlates tightly with the number of neurons firing at the corresponding electrode *in vivo* and *in vitro* (Plenz, 2012). Therefore, the critical branching parameter established for coherence potentials demonstrates a conservation law, specifically in which the initiating group size determines all group sizes that emerge within the coherence potential. This complements the finding that spike sequences at different locations within a coherence potential are similar (Plenz, 2012).

### OSCILLATION–SYNCHRONIZATION TRANSITION AND NEURONAL AVALANCHES: SIMULATIONS

Over the last decade, several models have explored these challenging relationships between oscillations, neuronal avalanches, coherence potentials and critical dynamics. The group of Linkenkaer-Hansen (Poil et al., 2008; Poil et al., 2012) demonstrated the emergence of avalanches with $\alpha \cong -3/2$ embedded in $\alpha$–oscillations (~12 Hz) in an E/I-balanced network model. They compared this to a critical parameter of $\kappa = 1$ and the emergence of long-range temporal correlations (see Fig. 12 below) in human MEG recordings demonstrating nested oscillations (Poil et al., 2012) [see also (Dalla Porta and Copelli, 2019)]. The co-emergence of oscillations and neuronal avalanches has been demonstrated in small systems to result from temporal correlations between large avalanches due to finite-size effects (Wang et al., 2016). It is an open question how such boundaries could be established in superficial cortical layers. When neuronal avalanches co-emerge with oscillations, neural networks achieve high cost-efficiency; that is, they balance their need for moderate synchronization with high information capacity (Yang et al., 2017). Recent models have combined system-wide synchronization and hysteresis, i.e. to support an oscillation cycle, with structural heterogeneity, i.e. to capture the variability observed in avalanches, in order to arrive at the co-emergence of oscillations and scale-invariant avalanche statistics (e.g. (di Santo et al., 2018)), while others have added an oscillating extinction rate to a continuous-time branching process using perturbative field theory (Pausch et al., 2019). Coherence potential-like activity and its potential computational advantages have been explored by the Gong group (Gong and van Leeuwen, 2009; Gong and Robinson, 2012; Chen and Gong, 2019). The many experimentally established dimensions of neuronal avalanches provide a rich testing ground to study the role of SOC in cortical information processing both experimentally and in network simulations. In the following sections, we will provide additional key experimental aspects of neuronal avalanche dynamics that go beyond size and synchronization scaling aspects.

### DEVELOPMENTAL SELF-ORGANIZATION OF ROBUST AVALANCHE DYNAMICS IN ORGANOTYPIC CORTEX CULTURES

The previous sections demonstrated the emergence of neuronal avalanches around the 2nd week postnatally in culture and in adult slices when tested in isolation. It is well understood that cortical development *in vivo* involves intrinsically maturing cellular properties and microcircuits in a complex interplay with structuring sensory inputs and motor outputs (Molnár et al., 2020). Many of these intrinsic embryonic and neonatal dynamics are found to arise autonomously in isolated cortex preparations (Allene and Cossart, 2010; Luhmann et al., 2016). Yet, so far only a few studies have reliably covered the time course of avalanche emergence during development over prolonged periods. In a first study of postnatal maturation of avalanches *in vitro*, Stewart and Plenz (2007) grew individual organotypic cortex cultures on a planar MEA in sterile chambers over many weeks (Fig. 8A). Spontaneous LFP activity emerged towards the beginning of the 2nd week postnatally with a typical bimodal distribution in cascade sizes (Fig. 8B – D) indicating a bias towards system-wide population bursts before the time of superficial layer maturation. During the end of the 2nd week, stable power laws in avalanche activity emerged particularly in those cultures that reached a high level of spontaneous activity and intermittent synchronized activity (Fig. 8B – F). Given the late development of superficial layers and the well-known preponderance of deep layer gap-junctions during the first postnatal week (Dupont et al., 2006), the initial bimodal distribution in cascade sizes might reflect system-wide deep-layer synchronization supported by extensive gap-junction coupling (Kandler and Katz, 1998), in turn potentially facilitated by transient hyperconnectivity which reduces towards the end of the second postnatal week *in vivo* (Meng et al., 2019). The ability of young cortex to express neuronal avalanches towards the end of the 2nd postnatal week was recently confirmed for superficial layers in young acute cortex slices (Bellay et al., 2021).

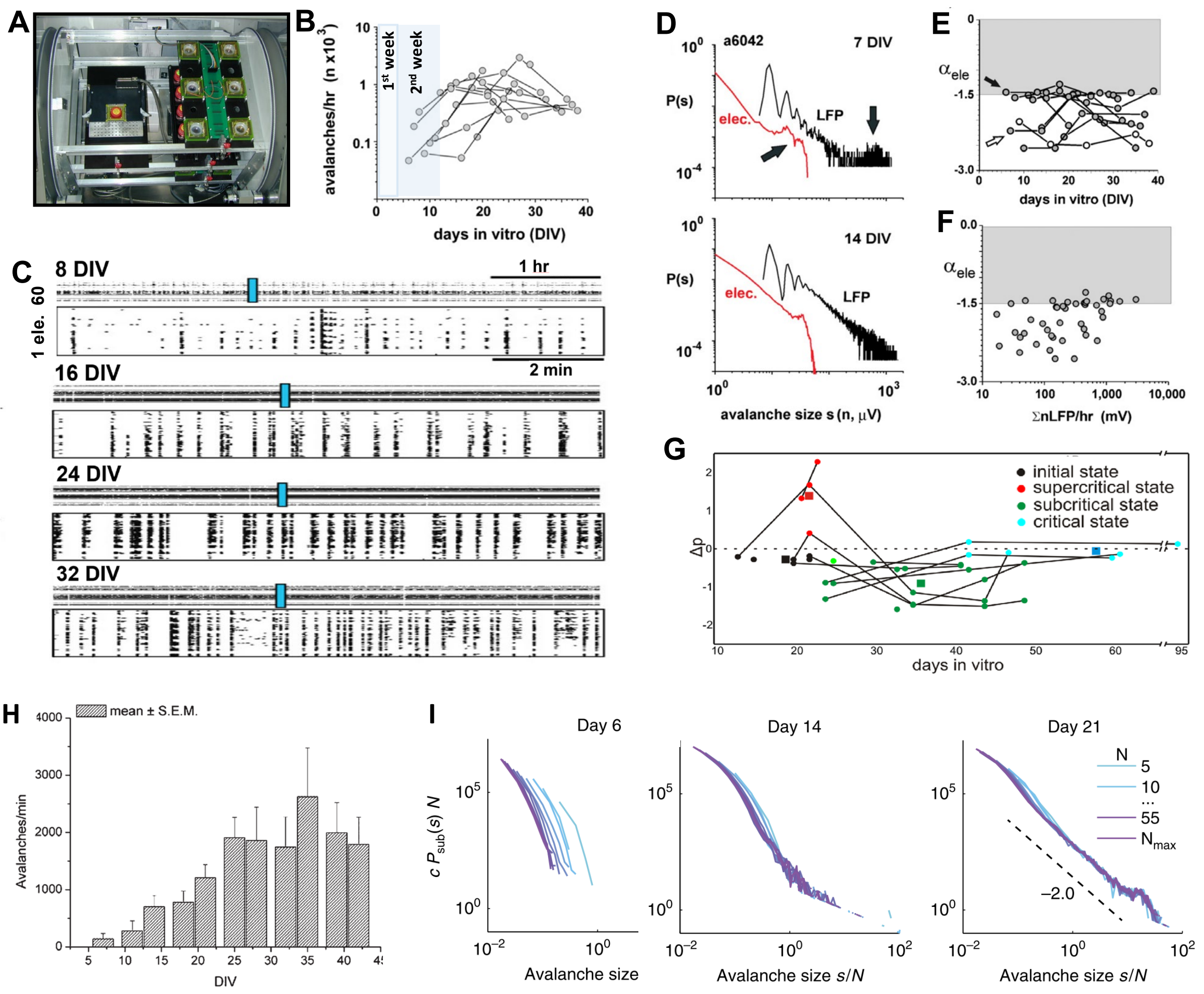

**Figure 8. Developmental time course for the self-organization of neuronal avalanches in isolated cortex preparations. A**, Overview picture of a custom-made incubator for long-term recordings of individual organotypic cortex cultures on a MEA in a chronic, sterile chamber with head stage (*left*) and off-recording storage racks (*right*). For details see (Plenz et al., 2011). [Reproduced with permission from (Pfeffer et al., 2004).] **B**, In organotypic cortex cultures, avalanches are absent during the 1$^{st}$ postnatal week *in vitro*, but increase in rate during the 2$^{nd}$ and 3$^{rd}$ postnatal week in line with *in vivo* maturation of superficial layers. *Highlighted periods*: equivalent postnatal week *in vivo* when cultures are taken from pups at postnatal day (PND) 1 – 2. **C**, Raster plots of spontaneous nLFPs increase in complexity from 1$^{st}$ to 5$^{th}$ week (*rows*) postnatally *in vitro*. *Top of each row*: 5 hr raster. *Bottom of each row*: Higher temporal resolution for periods indicated by the blue rectangle. **D**, Example of early (7 DIV) bimodal size distribution (*top*) and 2$^{nd}$ week power law size distribution (*bottom*) in a single organotypic cortex culture. Sizes defined as number of active electrodes (*red*) or summed nLFP (*black*). **E,** Most organotypic cortex cultures achieved $\alpha = -3/2$ within 2 – 3 weeks *in vitro*, except for one (*open circle*). Time course in avalanche size distribution slope α for individual organotypic cortex cultures. **F**, The emergence of slope $\alpha = -3/2$ correlates with an ~10x increase in LFP activity. [*B – F* modified and reproduced with permission from (Stewart and Plenz, 2007).] **G**, In dissociated cultures taken at PND 0 – 1, power laws tend to be reported after ~4 weeks in culture. [Reproduced under CC-NY license from (Tetzlaff et al., 2010).] **H**, Neuronal activity reaches steady state in dissociated cortex cultures after ~4 weeks *in vitro*. [Reproduced with permission from (Pasquale et al., 2008).] **I**, Transition of avalanche size distributions from exponential to bimodal in dissociated cortex cultures. [Reproduced under CC-NY license from (Levina and Priesemann, 2017).]

A second developmental study followed avalanche emergence in dissociated cortex cultures grown on MEAs, starting with neonatal cortex tissue around postnatal day (PND) 0 – 1 (Tetzlaff et al., 2010). This study described an initial bimodal size distribution as well, characterized as 'supercritical', followed by a pronounced 'subcriticality' and eventually, after more than five weeks in culture, a 'critical' condition characterized by stable power laws in size distribution (Fig. 8G). While both organotypic and dissociated cultures capture an initial bimodal activity state, the developmental time course in dissociated cultures appears to be delayed by more than three weeks with respect to the build-up of neuronal activity (Fig. 8H) and power law formation when compared to *in vivo* (Gireesh and Plenz, 2008). Recently, Levina and Priesemann (2017) showed that the bimodal distribution in avalanche sizes is maintained in dissociated cultures over long periods, questioning the robustness of power laws identified by previous studies for that system (Fig. 8I; see also below).

These developmental studies of cortical tissue in isolation suggest distinct differences in neuronal avalanche emergence between organotypic and dissociated cortex cultures, with the latter demonstrating a delayed maturation time course compared to *in vivo* and a tendency of bimodal size distributions. In contrast, avalanche emergence in organotypic cortex cultures matches that of the *in vivo* development with respect to layer location and robustness in power law scaling.

## SEPARATION OF TIME SCALES AND HOMEOSTATIC REGULATION OF NEURONAL AVALANCHES

A separation of time scales, in which the time course of driving the system is slow enough as to not interfere with the fast avalanching process itself, is of essence in some models of SOC (e.g., (Hesse and Gross, 2014)). This concept was tested in organotypic cortex cultures grown individually in sterile, closed chambers while the chamber is tilted periodically. This approach periodically submerges the culture in liquid culture medium ('feeding') followed by exposure to normal air ('breathing') and slowly drives the system through concommittant, large changes in neuronal population activity (Fig. 9A, *top*; (Plenz, 2012)). The resulting avalanche size distributions were power law distributed despite strong common, external triggers from the change in environmental condition (Fig. 9A; *bottom*). In a second series of experiments, the well established effect of rebound activity and rebound bursts after prolonged periods of suppression in excitatory synaptic transmission (Fig. 9B, C; (Corner et al., 2002; Turrigiano and Nelson, 2004)) was used to study the robustness of avalanche dynamics. Excitatory synaptic transmission was mildly reduced in organotypic cultures by adding a low amount of the fast glutamate receptor antagonist, DNQX, to the culture medium for 24 hr. This reduction in excitatory transmission steepened the distribution in cascade sizes. Importantly, after removing the 'brake' on excitatory transmission, cascade size distributions rapidly became bimodal with an initial steep slope close to -2, but autonomously recovered within 24 h towards the power law distribution with a slope of -3/2 observed prior to the perturbation (Fig. 9D; (Plenz, 2012)). These experiments demonstrate homeostatic regulation of avalanche dynamics from a supercritical to a critical state in the absence of any structuring external inputs. A recent study by Ma et al. (2019) demonstrated recovery to power law distributed avalanches during monocular deprivation *in vivo* over the course of several days, suggesting that recovery can be initiated from the subcritical phase as well.

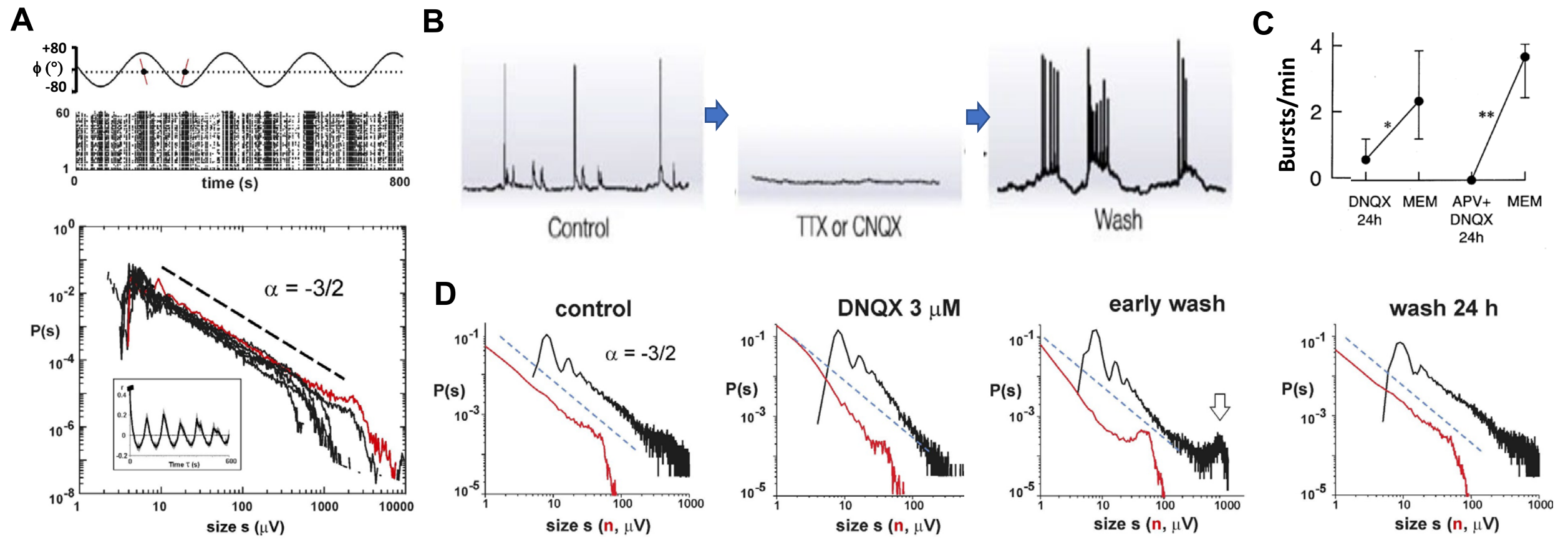

**Figure 9. Separation of time scales and homeostatic regulation of neuronal avalanches *in vitro*.** **A**, Power law in avalanche sizes is preserved during externally induced slow changes in neuronal activity. *Top*: Periodic tilting (ϕ) of the sealed, sterile chamber, which contains the organotypic culture grown on the MEA, alternates culture exposure to feeding medium or air. This slowly drives the network through corresponding large changes in nLFP activity visualized in the corresponding raster plot. *Bottom*: Power law in avalanche sizes for $n$ = 7 cultures under conditions of periodic tilting. *Red*: Distribution from single culture shown on top. *Inset*: Average autocorrelation of population activity demonstrating strong correlations from periodic tilting. **B**, Intracellular recording demonstrating rebound hyperactivity when neuronal activity is suppressed in dissociated cortex cultures. [Reproduced with permission from (Turrigiano and Nelson, 2004).] **C**, Increase in rebound population bursts after 24 hr of excitatory glutamate receptor suppression in organotypic cortex cultures. [Reproduced with permission from (Corner et al., 2002).] **D**, Homeostatic regulation of avalanches in organotypic cortex cultures. *Control*: Size distribution before manipulation. *DNQX*: Size distribution during 24 hr of reducing excitatory synaptic transmission. *Early wash*: Bimodal size distribution demonstrating hyerexcitability after wash out of DNQX. *Wash 24 hr*: Homeostatic return to avalanche dynamics before pharmacolocial manipulation. Distributions based on 1 hr recordings each from a single culture. [*A*, *D* reproduced with permission from (Plenz, 2012).]

### Size-dependent nesting in the temporal self-organization of neuronal avalanches

Population activity that spontaneously forms in isolated cortex preparations has been typically described as intermittent bursts of variable length, as well as variable intensity, pauses in between (*cf.* Fig. 8C). When analyzing the summed population activity of avalanche activity more closely, the picture of 'avalanches within avalanches' readily emerges (Fig. 10A, B), which dominates periodically driven cultures as well (Plenz, 2012). Based on the observation of an avalanche, the time to wait before observing a future avalanche is known as the waiting time and was found to reflect the characteristic time scales of θ/γ–oscillations and up-states (Fig. 10C; (Lombardi et al., 2014)). This was true for avalanches independent of minimal size and with strong dependence on the E/I balance (Lombardi et al., 2012). By calculating conditional probabilities, Lombardi et al. (2016) obtained precise functions capturing the nesting of avalanches with respect to size and time to the next avalanche. It was generally found that there is a high probability that the next avalanche will be smaller than the currently observed avalanche. This finding was robust to a large range of sizes and time windows of observation as well for periodically driven activity (Fig. 10D; *left*, *middle*). Importantly, this relationship reverses when reducing inhibition, more specifically the network becomes hyperexcitable at which point future 'avalanches' are likely to be larger than the currently observed activity (Fig. 10D; *right*; cf. also Fig. 7D). These experimental findings add an important dimension to the discussion of hyperexcitable network activity beyond the finding of bimodal size distributions.

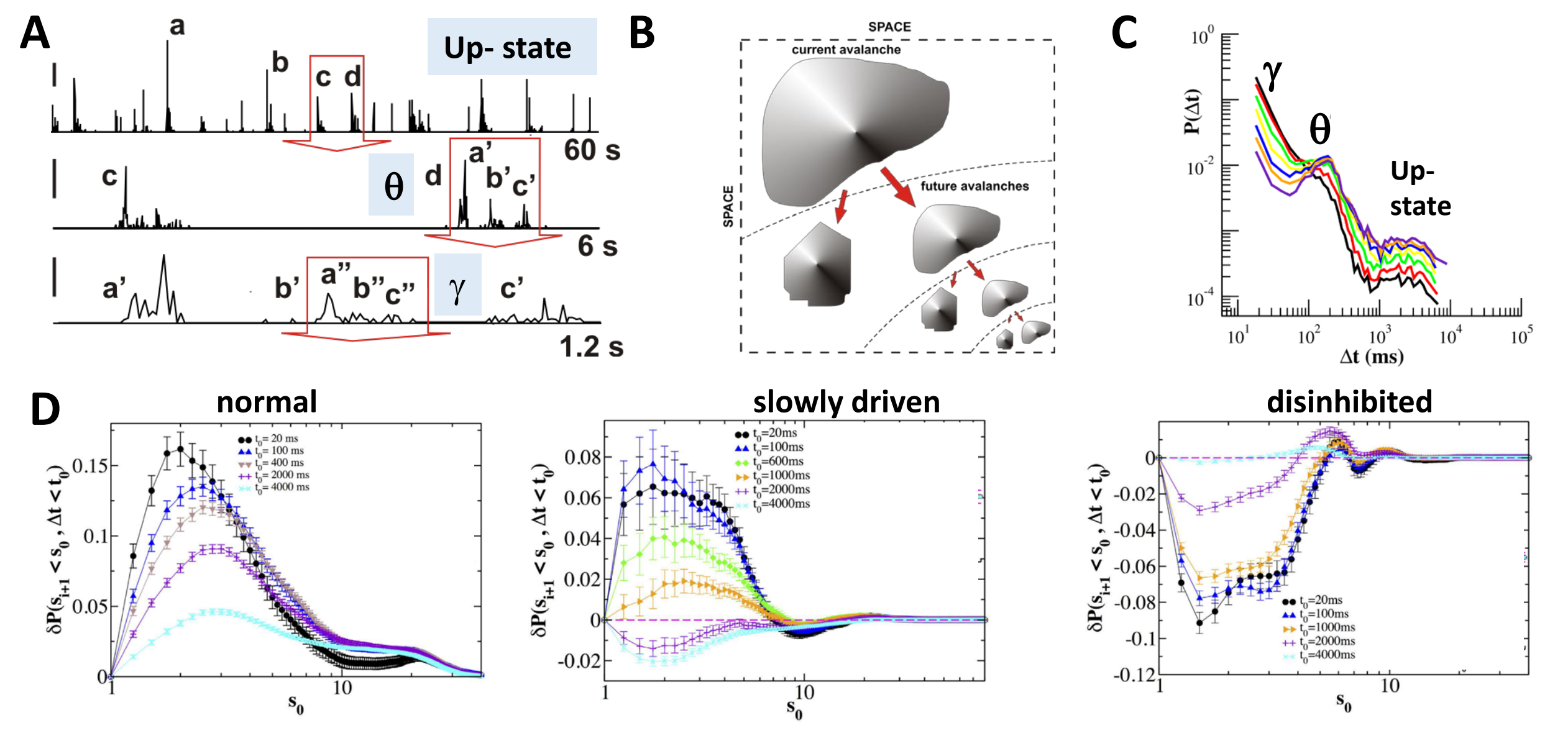

**Figure 10. Size-dependent nesting in the temporal self-organization of neuronal avalanches. A**, Time course of integrated avalanche activity at three different resolutions. Note the presence of intermittent bursts of successively smaller size (e.g., (*a, b, c, d*) that are recursively embedded at smaller time scales (e.g. *d* -> (*a', b', c'*); *b'* -> *(a'', b'', c'')*). *Red arrows* indicate zoomed-in period from larger time scale (single organotypic culture; slowly driven; *cf*. Fig. 9A). Labels place each time scale within the corresponding up-state and nested oscillation regimes. **B**, Sketch of size dependent nesting of avalanches begetting future avalanches. [*A, B* Modified and reproduced with permission from (Plenz, 2012).] **C**, Quiet time distribution of avalanches revealing power law decay with θ–oscillation peak and indicated γ–oscillation and up-state regime (organotypic cortex cultures) [Modified and reproduced under CC-BY license from Lombardi et al. 2014).] **D**, Quantification of the dependency in pre-avalanche size and successive quiet time. Avalanches beget future, smaller avalanches (*left*) absent driving, and during periodic, slowly driven condition (*middle*). This relationship is reversed in disinhibited cultures (*right*). Y-axis displays the probability that the next avalanche within observation window $t_0$ is smaller given size $s_0$ of the current avalanche. [Reproduced under CC-BY license from (Lombardi et al., 2016).]

## CONTROL PARAMETERS IDENTIFIED IN THE REGULATION OF NEURONAL AVALANCHES

The core requirement for SOC is the ability for the system to adjust a control parameter, which allows the system to reside near the critical point (Hesse and Gross, 2014; Chialvo et al., 2020). Given the complexity of cortical microcircuits regarding neurotransmitter categories (excitatory, inhibitory), neuromodulators (e.g., dopamine, acetylcholine, serotonin) and brain states (e.g., wakefulness, sleep, attention), there could be many control parameters that are able to tune cortical networks towards or away from criticality, yet few have been experimentally examined so far. Of common focus, the E/I balance establishes an important control parameter, first demonstrated for avalanches in organotypic cortex cultures (Beggs and Plenz, 2003). Specifically, reducing fast inhibitory synaptic transmission non-selectively by pharmacological means, rapidly destroys the power law in LFP-based avalanches and causes bimodal distributed cascade sizes (*cf.* Fig. 5E). Similar results have been obtained in dissociated cultures, in which a power law distribution in avalanches changed to a bimodal distribution when inhibition was blocked (e.g., (Pasquale et al., 2008)). In more detailed follow up studies, a reduction in fast synaptic inhibition or in fast and slow synaptic excitation changes the dynamics from avalanches to a 'supercritical' or 'subcritical'-like condition (Shew et al., 2009; Shew et al., 2011; Yang et al., 2012; Shew and Plenz, 2013). These studies, by quantifying the distance of bimodal or exponential distributions from a power law, demonstrated that numerous network parameters are maximized at the E/I balance at relatively low level of synchronizations, where avalanche dynamics reigns (Fig. 11).

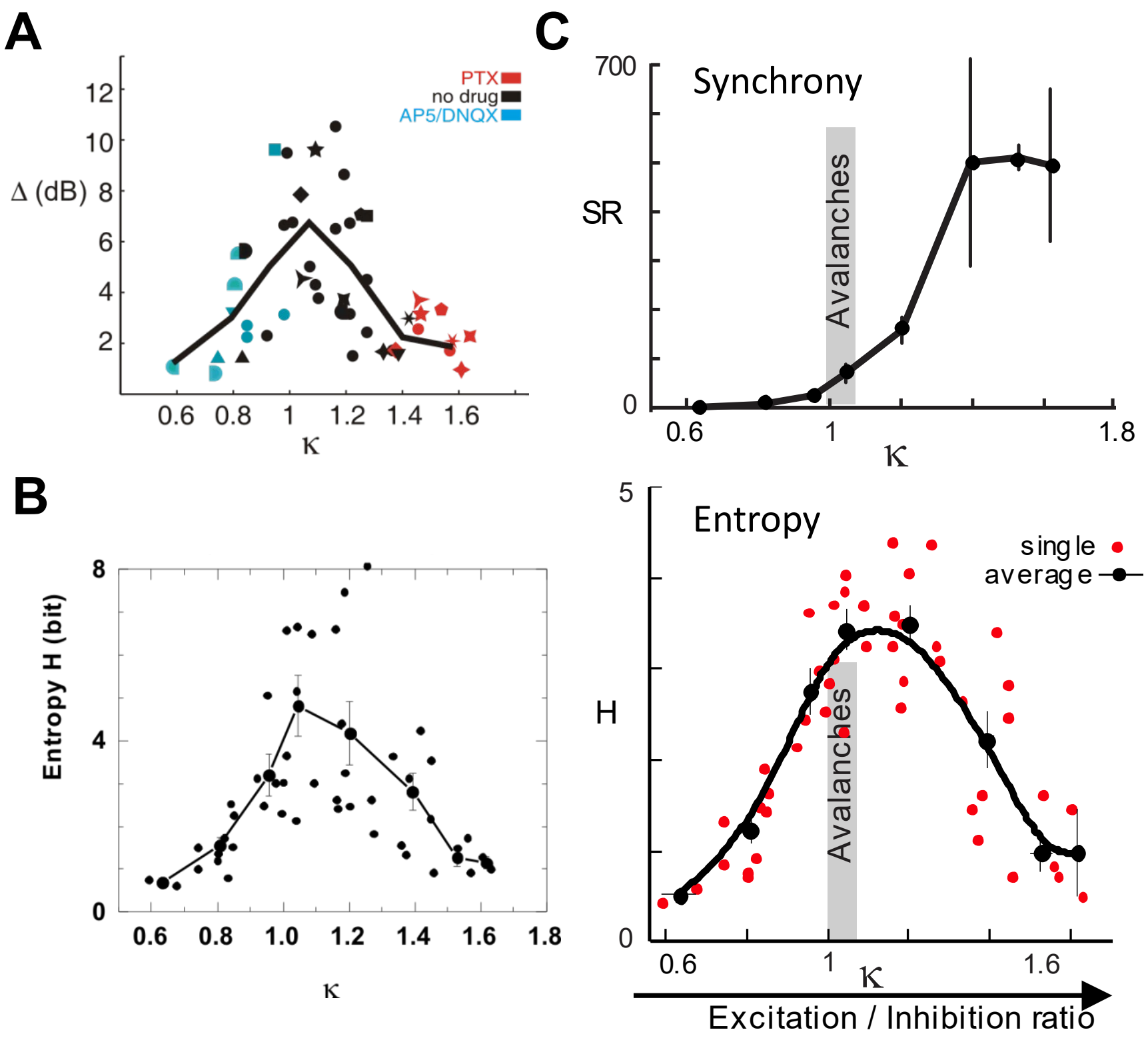

**Figure 11. The E/I balance is a control parameter for the emergence of avalanches. A**, The dynamic range Δ is maximized when avalanche size distributions are closest to a power law. The parameter κ quantifies the Kolmogorov-Smirnov deviation at 10 equidistant steps of the actual cumulative size distribution from that of a power law (Shew et al., 2009; Shew et al., 2011). When κ = 1, the distribution is a power law, whereas κ >1 for a bimodal distribution and κ < 1 for an exponential distribution. **B**, Information capacity is maximized close to κ = 1. **C**, Synchronization exhibits a phase transition at κ = 1 (*top)* where the entropy of synchronization is maximal (*bottom*) and total synchronization is still an order of magnitude lower than that in the hyperexcitable regime for κ > 1. *A – C* are derived from LFP avalanches measured in organotypic cortex cultures with κ changed pharmacologically as indicated in *A*. [Reproduced from (Shew et al., 2009; Shew et al., 2011) and (Yang et al., 2012) respectively. Copyright 2009, 2011, 2012 Society for Neuroscience.]

The neuromodulator dopamine has been identified as a second control parameter for the regulation of neuronal avalanches in prefrontal cortex (Fig. 12). Dopamine is crucial for working memory performance, which in turn requires prefrontal cortex functioning (Durstewitz and Seamans, 2008; Arnsten, 2011). Acute prefrontal cortex slices taken from adult rats, exposed to a moderate external excitatory drive, rapidly respond to the presence of dopamine with the emergence of avalanche activity in superficial cortex layers (Fig. 12A, B; (Stewart and Plenz, 2006)). At intermediate levels, but not low or high levels of dopamine, nLFPs formed a power law in avalanche sizes with slope of -3/2 (Fig. 12C). The activity was selective for the dopamine $D_1$-receptor and required NMDA-glutamate receptor stimulation, thus matching the pharmacological inverted-U profile reported for working memory performance in prefrontal cortex (Cai and Arnsten, 1997). Analysis of the intracellular membrane potential in individual pyramidal neurons in the acute slice, as well as extracellular single-unit analysis *in vivo*, demonstrated that even large LFP avalanches engage individual pyramidal neurons selectively and this selectivity breaks down when inhibition is reduced (Bellay et al., 2021). These results taken together suggest that the control parameter dopamine maximizes the spatial extent and occurrence frequency of system-wide avalanches formed by selective activation of distributed pyramidal neurons in the network.

The high sensitivity of the power law to the reported control parameters suggests that thresholding of the LFP is unlikely to play a major role in the origin of scale-invariant avalanches. The LFP is a continuous time-varying signal, for which avalanche processing requires a threshold operation to convert this signal into point process-like data. Such thresholding preserves essential avalanche information in a discretized spatiotemporal raster, e.g., as shown for human avalanches in the fMRI (Tagliazucchi et al., 2012). Yet, thresholding is a non-linear operation and can affect scaling regimes particularly in the temporal domain (Francesc et al., 2015; Villegas et al., 2019). On the other hand, if thresholding were the underlying cause to observe avalanches, one would expect power law characteristics in the observed dynamics to be robust to relatively mild pharmacological manipulation, which is not the case.

Changes in network connectivity based on local plasticity rules have been demonstrated to establish SOC in models (Bornholdt and Rohlf, 2000), suggesting that plasticity could function as a control parameter (e.g. (de Arcangelis and Herrmann, 2010; Michiels van Kessenich et al., 2018)). Since network connectivity was found to support avalanche dynamics in dissociated cultures, it could be considered as a control parameter as well (Massobrio et al., 2015). On the other hand, measurements in organotypic cortex cultures and in nonhuman primates *in vivo* demonstrate avalanches establish integrative network architectures that are robust to certain plastic changes (Pajevic and Plenz, 2009; 2012; Miller et al., 2021). Of note, *in vivo* studies have shown avalanches to be exquisitely sensitive to the sleep/wakefulness transition (Ribeiro et al., 2010; Scott et al., 2014; Bellay et al., 2015; Fagerholm et al., 2016; Ribeiro et al., 2016) suggesting sleep (Meisel et al., 2013; Meisel et al., 2017) and sleep-arousal transitions (Lombardi et al., 2020) as a behavioral state control parameter.

## LACK OF SCALE-INVARIANT LFP AVALANCHES IN DEEP LAYERS

Results summarized here were based on LFP recordings taken from high-density arrays oriented in a specific manner with respect to the underlying cortical column. The planar projection of the array was aligned in such a way where propagation of activity in all layers of cortex can be monitored. Even under those carefully chosen projection conditions, deep layer LFP activity was either strikingly absent, e.g., during spontaneous avalanche emergence in organotypic cortex cultures (*cf.* Fig. 3) or during external glutamate mediated depolarization, which induces avalanche activity in superficial layers in the acute cortex slice (Fig. 13). The absence of LFP-based avalanches in deep layers *in vitro* could have various causes. First, deep layers could mature incompletely in organotypic cultures preparations, e.g., due to lack of subcortical inputs from thalamus or lack of subcortical targets. However, this argument does not apply to the acute, mature cortex slice. Second, deep layers might require the presence of neuromodulators such as acetylcholine and neurotensin, often provided by brain regions outside the cortex, which regulate the amount of bursting in deep layer pyramidal neurons (McCormick and Prince, 1987; Case et al., 2016). However, even in the awake nonhuman primate, the LFP activity in deep layers does not establish power laws even when avalanche activity propagates simultaneously in superficial layers (Fig. S4 of (Petermann et al., 2009)). Third, avalanches in deep layers could be composed of spatially distributed neurons that are difficult to track in the LFP. However, local cortical connectivity favors connections between nearby pyramidal neurons (Braitenberg and Schüz, 1991), such that avalanche activity would be expected to sum in the LFP. Taken together, these arguments suggest that deep layers might not be able to support scale-invariant avalanche dynamics in general. Even advanced recording techniques *in vivo* in the awake rodent demonstrate absence of avalanches in deep layers. Using 2-photon imaging *in vivo*, power laws in spike-based avalanches were identified in cortical layer 2/3 and layer 4 (Bowen et al., 2019), but seem to be absent in deep layers (Ma et al., 2020).

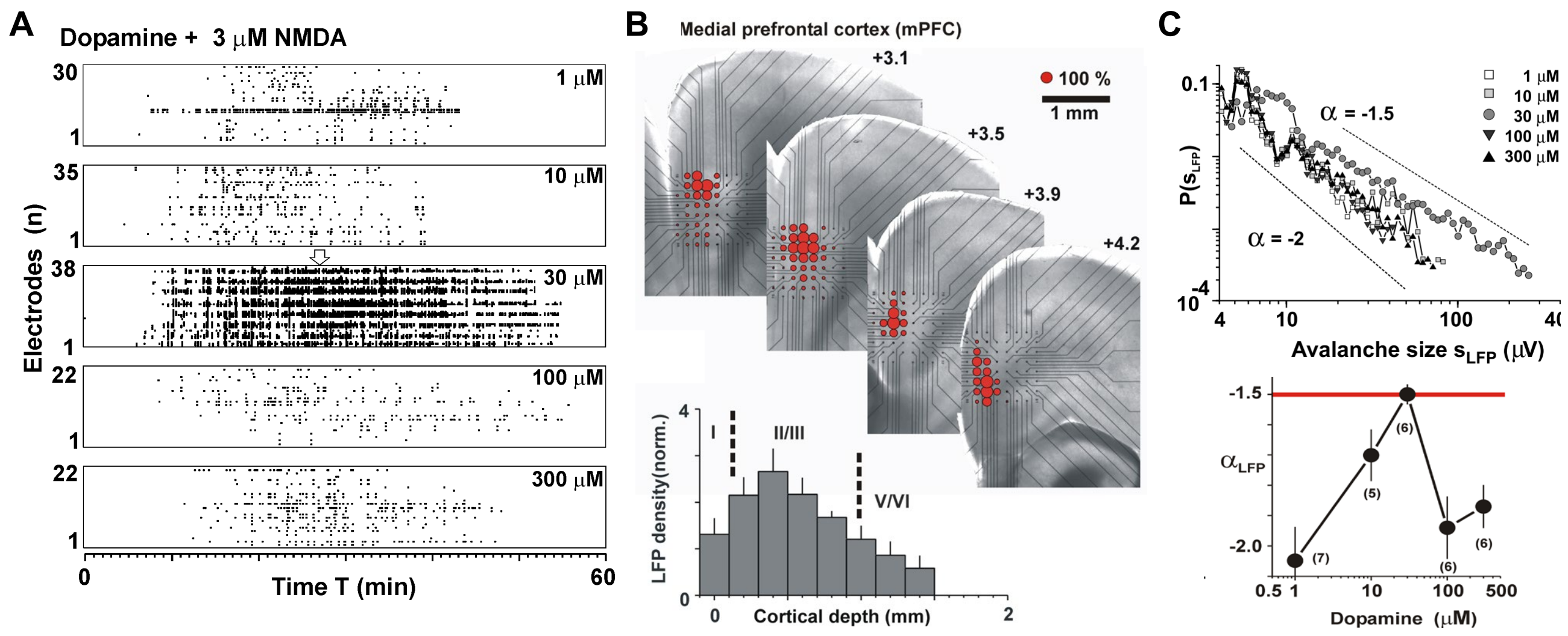

**Figure 12. Dopamine is a control parameter for the emergence of neuronal avalanches in superficial layers of cortex. A**, Externally-driven acute cortex slice using weak excitatory drive (3 μM NMDA) maximizes avalanche activity at intermediate dopamine concentrations. Raster plots of nLFPs from different slices for increasing dopamine concentrations (*top to bottom*). **B**, Externally-driven neuronal avalanches emerge in superficial layers 2/3 mainly. Average nLFP density (*red circle size*) projected onto light microscopic image of the MEA with corresponding acute coronal slice from medial prefrontal cortex. *Lower left*: nLFP density projected along cortical layers demonstrating that most avalanche activity is induced in L2/3. **C**, Avalanche activity exhibits a power law slope of -3/2 at moderate dopamine concentration. Size distributions (*top*) and corresponding slopes (*bottom*) as a function of dopamine concentration. [Reprinted with permission from (Stewart and Plenz, 2006). Copyright 2006 Society for Neuroscience.]

### The nLFP is the avalanche – A local reconstruction from spike avalanches using 2-photon imaging

In the LFP, the structural and dynamical heterogeneity of the network is summed to form a local point source, which does not allow for the identification of the network elements contributing to the LFP (Buzsáki et al., 2012). While many experimental findings on avalanches have utilized spatially expansive MEAs, scale-invariance predicts that avalanche dynamics should be observable even within the local neighborhood of a single electrode as the spatial resolution increases. This in turn should allow for a more detailed analysis of the underlying network components contributing to scale-invariance. In this scenario, the nLFP amplitude should reflect the local neuronal group activity governed by avalanche dynamics (Fig. 13A). Accordingly, it was found that the nLFP amplitude distribution at the single-electrode level approximates a power law with slope of -3/2, which is destroyed when pharmacologically changing the E/I balance (Fig. 13B; (Plenz, 2012)).

The notion that locally summed activity of neuronal group *firing* constitutes avalanche dynamics was first demonstrated directly with 2-photon imaging using the genetically encoded calcium indicator (GECI) YC2.60, which exhibits single spike sensitivity (Yamada et al., 2011; Bellay et al., 2015). The indicator was selectively expressed in pyramidal neurons from superficial layers in organotypic cortex cultures using electroporation (Fig. 13C; (Bellay et al., 2015)). When the coordinated firing in groups of pyramidal neurons was studied, it was found that the highly irregular firing of pyramidal neurons during ongoing spontaneous activity exhibits clear avalanche signatures. We note that these power laws are robust at temporal scale of $\Delta t$ = 33 ms (i.e. at imaging frame rate of 30 Hz) and their slope $\alpha$ is shallower than -3/2 as predicted from LFP avalanche analysis (cf. Fig. 5B, G). Importantly, the power law in avalanche sizes was transformed to a bimodal distribution only after pharmacologically reducing inhibition at which the typical hyperexcitable phenotype of an initial steep slope close to -2 and an overabundance of system wide cascades robustly presents (Fig. 13D – F). We note in passing that a hyper-synchronized phenotype in the size distribution also emerges when mildly reducing excitatory transmission (cf. Fig. 13B, F; *right* 'Disfacilitated'). Such a reduction increases spontaneous synchronization in the network due to an overall increase in synaptic transmission efficacy when the global rate of activity drops (e.g. (O'Donovan, 1999)). Taken together, these results demonstrate that the nLFP reflects local avalanche activity and should not be equated to single spikes. In this context, the threshold, $\lambda$, applied to extract nLFPs is similar to the population threshold applied to summed spiking activity in identifying neuronal avalanches in network models (e.g. (Poil et al., 2012)). These findings identify pyramidal neuron activity in superficial cortex layers to carry signatures associated with the organization of avalanches, which since then has been confirmed *in vivo* (Bellay et al., 2015; Karimipanah et al., 2017; Bowen et al., 2019; Ma et al., 2020). The relationship between firing statistics of single neurons and critical exponents in avalanche dynamics has been a major research topic in neural network dynamics (e.g. (Buice and Cowan, 2007; Benayoun et al., 2010; Buice et al., 2010; Liang et al., 2020)).

### Not all avalanches are SOC avalanches – The prevalence of local and system-wide population events in Dissociated neuronal cultures

Dissociated cultures (Dichter, 1978) have been used for decades to study the autonomous development in structure and dynamics of cortical microcircuits. As a complementary approach to organotypic cultures (Gähwiler et al., 1997), dissociated neuronal cultures are prepared from cortical tissue typically taken from an embryo at embryonic day 15 – 18 (that is 3 – 6 days before birth). The tissue is then mechanically and enzymatically disintegrated, the remaining neuronal cell bodies and pre-cursor cells are re-seeded on a glia-feeder layer, and grown for up to several months *in vitro* (Dichter, 1978). Dissociated cultures appeal by focusing on the *de novo* formation of neuronal connections, yet they require careful attention to the design of glia-feeder layers and the culture medium composition. They lack cortical layers and a clear classification of pyramidal neurons and interneurons into functional subtypes, which, in contrast, are both well established *in vivo* and organotypic cortex cultures during various developmental stages (see introductory sections). Synchronized bursting has been the hallmark of the developing population activity in dissociated cultures grown on MEAs (Maeda et al., 1995). Despite the apparent simplicity of this culture system, when systematically studied using a large number of cultures grown on MEAs over many weeks, highly variable outcomes in neuronal synchronization have been documented that depend on plating density, which affects the number of neurons per area and developmental trajectory (Wagenaar et al., 2006). Accordingly, the application of avalanche analysis to these synchronized bursts has yielded heterogenous outcomes across and within studies (Fig. 14). Nevertheless, a consistent finding emerges from these studies, which deviates from results reported for LFP- and spike-based avalanches in organotypic cortex cultures, as detailed below.

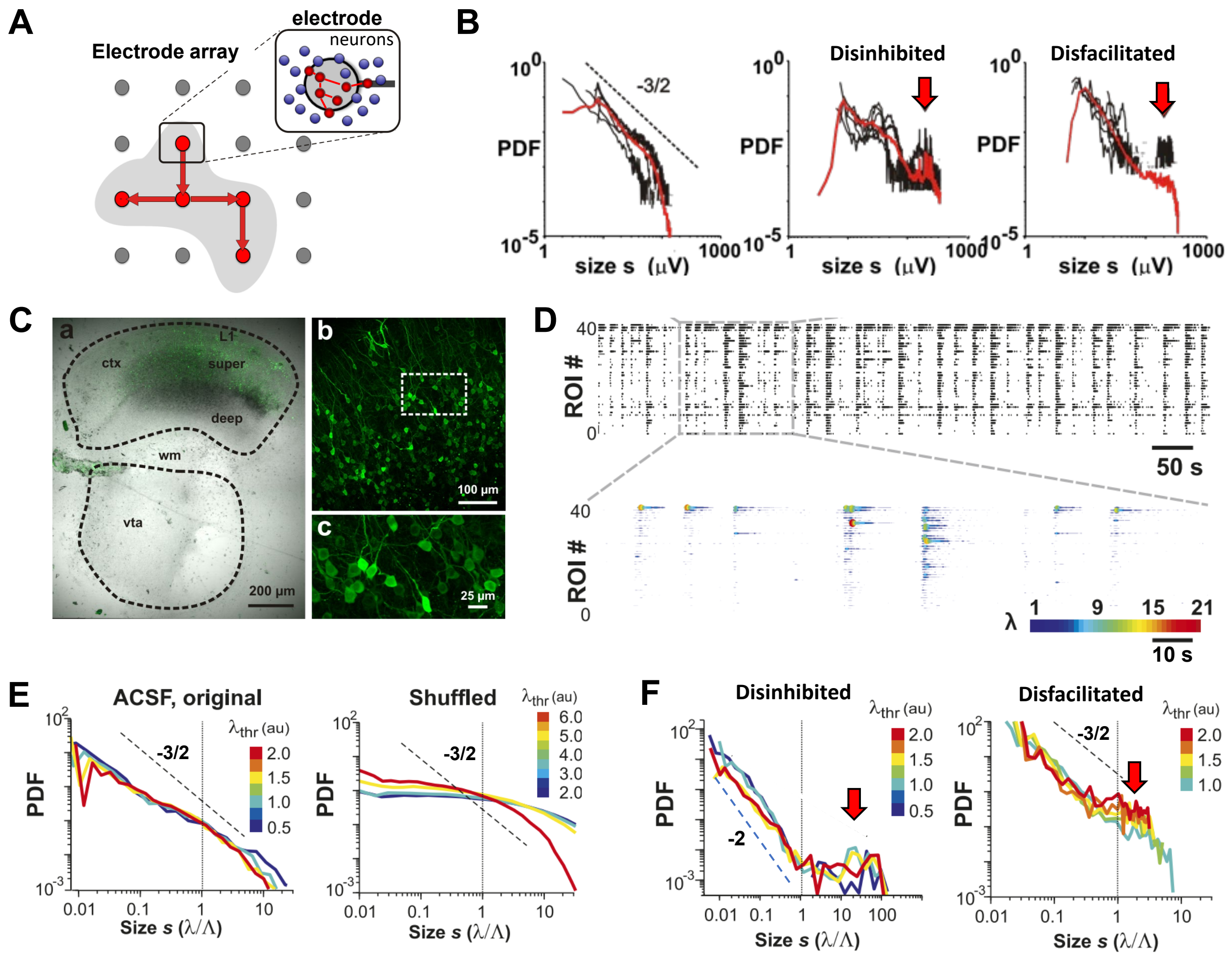

**Figure 13. Experimental transition from single nLFP avalanches to spike avalanches in organotypic cortex cultures using 2-photon imaging (2PI). A**, Sketch of spatial transformation of propagated nLFPs on the electrode array to propagated spike activity in local neuronal groups (*circles*; *red*: spiking; *blue*: quiet) within the neighborhood of a single electrode (*zoom*). **B**, Distribution of nLFP amplitudes at single electrodes (*red*: average; *black*: example single electrodes from organotypic culture). Power law-like distributions change into bimodal distributions when reducing inhibition (*disinhibited*) or excitation (*disfacilitated*). [Reproduced with permission from (Plenz, 2012).] **C** – **F**, Reconstruction of spike avalanches in organotypic cortex culture. **C**, Single organotypic co-culture of cortex (*ctx*) and ventral tegmental area (*VTA*) taken at PND 0 – 1 and grown for ~3 weeks. Electroporation of the embryo at E16.5 leads to expression of the genetically encoded calcium indicator YC2.6 in pyramidal neurons from superficial cortex layers (*b* and *c* are successive zooms from *a*). *Broken lines*: Tissue borders. *wm*: White matter. **D**, Raster of spontaneous spike density monitored with 2PI and obtained through deconvolution (n = 40 pyramidal neurons). *Top*: Binarized raster. *Bottom*: Temporally expanded raster segmented with color coded spike intensity $\lambda$ for each neuron. **E**, Spontaneous neuronal activity reveals power laws in spike avalanches that are robust to the threshold $\lambda_{thr}$ applied at the single neuron level (*left*). Temporal shuffling of spike activity abolishes the power law in avalanche sizes (*right*). **F**, Mild disinhibition changes the power law to a bimodal distribution with initial steep slope of -2 and system-wide population events (*red arrow; cf. Fig. 5E*). System-wide events also become prominent when mildly reducing excitation (*right*). [Modified for *C* and *C* – *F* reproduced under CC0 license from (Bellay et al., 2015).]

Pasquale et al. (2008) were first to report spike-avalanches in six dissociated cultures with size distribution of either exponential, bimodal or power law form. Two out of six cultures displayed the power law in avalanche size, however, only at sub-millisecond temporal resolution *Δt*. In fact, increasing *Δt* to 1 ms steepened the initial slope and rapidly uncovered a bimodal distribution, explained in their model as explosive growth introduced by neuronal hubs (Fig. 14A). Using neonatal tissue right after birth, Tetzlaff et al. (2010) tracked spike avalanche distributions during development and found an initial slope close to -2 for mature cultures and an increase in bimodality with increasing bin size (Fig. 14B). Similar findings were presented by Levina and Priesemann (2017) using dissociated cultures prepared from E18 tissue and grown for ~3 weeks. Spike avalanches revealed a size distribution with a steep slope close to -2 and a preference for large avalanches. Again, an increase in *Δt* steepened the initial slope further and disproportionately increased the probability for large activity events (Fig. 14C). Even when examining only a small range of bin width, reported results for dissociated culture experiments are more in line with expectations for a hyperexcitable system. Yada et al. (2017) tracked the development of spike avalanches in six cultures and reported a bimodal form and an initial steep slope close to -2 that was robust to modest changes in *Δt* (Fig. 14D). Similarly, robust bimodality and initial steep slope was reported by Haukat and Thivierge (2016) (Fig. 14E).

These experimental findings are in contrast to what has been reported for sizes of LFP and spike avalanches in organotypic cultures, where an increase in *Δt* leads to a more shallow slope in the size distribution and bimodality only arises when the system is made hyperexcitable (*cf.* Fig. 5; Fig. 13). The notion that dissociated cultures might be characterized better by hyperexcitable dynamics is in line with a recent finding that increasing inhibition by adding a GABA-agonist reduces bimodal size distributions in dissociated cultures, bringing them closer to a power law (Heiney et al., 2019). It is striking that, in some dissociated cultures, power laws were found mainly at sub-millisecond *Δt* (e.g. (Pasquale et al., 2008)). Given that synaptic integration times between neurons are on the order of 2 – 3 ms, a sub-millisecond temporal integration window will prematurely terminate the tracking of propagated activity, thereby randomly partitioning synchronous activity and potentially creating heavy-tail statistics in cascade sizes. Accordingly, at temporal resolutions > 3 – 4 ms, which overlaps with synaptic integration times, a strongly bimodal

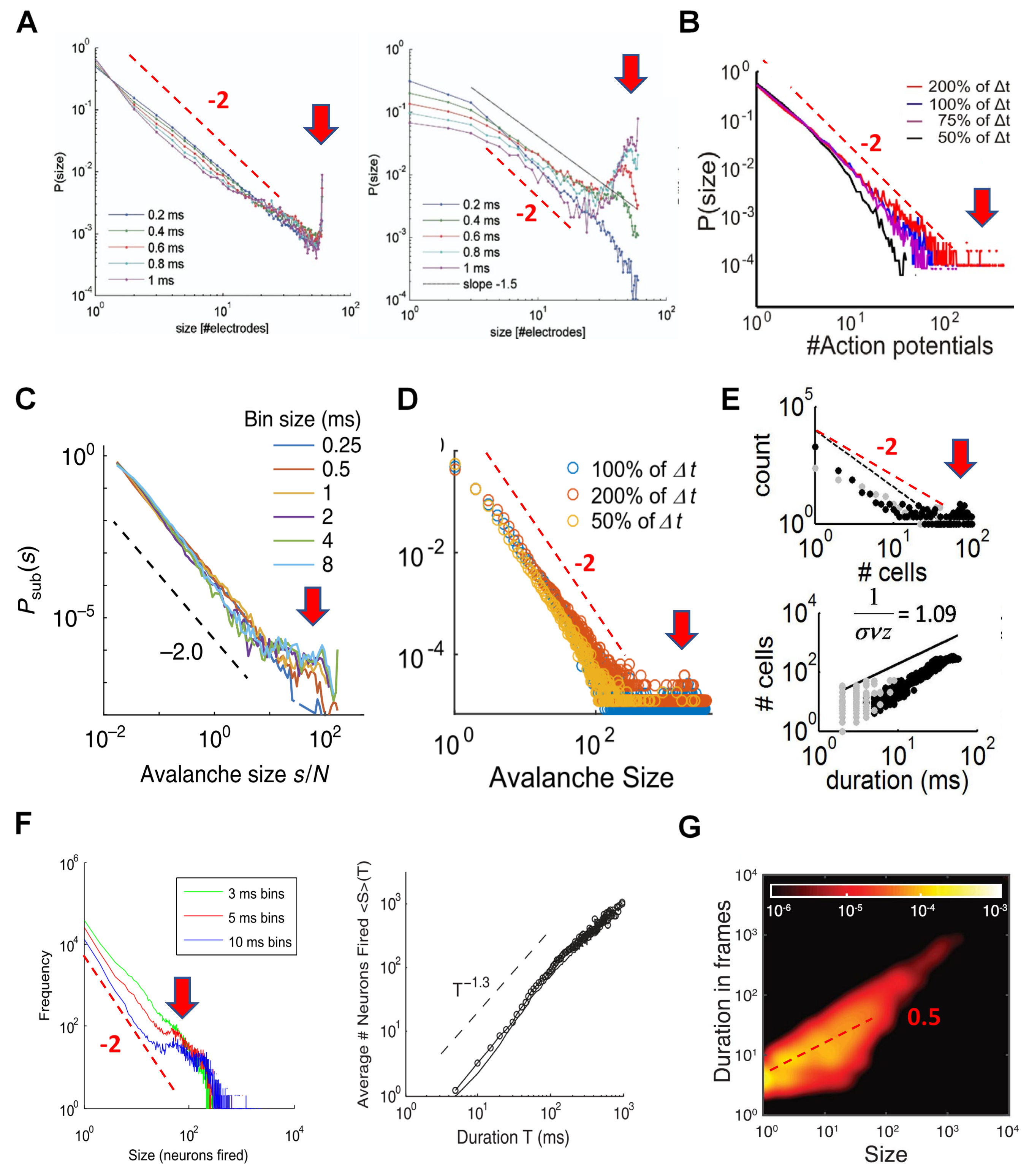

**Figure 14. Spike 'avalanches' in dissociated cultures display scaling characteristic of hyperexcitable dynamics.** **A**, Spike avalanches in dissociated cultures exhibit a steeper initial slope and become more bimodal with increase in *Δt*. *Left*: Cultures; *Right*: Model. [Reproduced with permission from (Pasquale et al., 2008).] **B**, With increase in *Δt,* spike avalanches in dissociated culture change from exponential to bimodal size distribution with steep initial slope close to -2. [Reproduced under CC-NY license from (Tetzlaff et al., 2010).] **C**, Initial slope of -2 and bimodality with increase in *Δt* in mature dissociated

population dynamics of local, non-propagated activity and global, propagated activity is revealed in these systems. For comparison, temporal resolutions studied in organotypic cultures ranged up to 16 ms (LFP avalanches; Fig. 5B,G) and 33 ms (spike avalanches; imaging frame rate of 30 Hz; Fig. 13). Despite these long integration windows, bimodal distributions were absent, unless the system was made hyperexcitable. These integration windows are about an order of magnitude longer than those at which dissociated cultures show clear bimodal dynamics.

We suggest that the bimodal size distribution in dissociated cultures reflects a predominance of local activity and system-wide propagated activity. Such heterogenous dynamics can arise from several scenarios. The bimodality could reflect different structural networks, potentially including different cell types, that mature in dissociated cultures. For example, Orlandi et al. (2013) applied an avalanche algorithm to neuronal activity tracked with intracellular calcium imaging in dissociated cultures grown from embryos for up to three weeks. They separated functional networks of 'background' avalanches that established a cascade size distribution steeper than -2 from system-wide avalanches. Alternatively, dissociated cultures could establish a homogenous neural network in which bimodality arises from a discontinuous, 1$^{st}$ order phase transition. In this latter case, neuronal activity either remains local and small or, alternatively, propagated and system wide. Increasing *Δt* in such a system more robustly collects activity into system-wide events, concomitantly reducing smaller-sized events steepening the initial slope in size distributions as observed for dissociated cultures. In fact, the simulation of population activity in dissociated cultures using a 1$^{st}$ order phase transition has a long tradition. These dynamics were captured in early models featuring a 1$^{st}$ order phase transition of all-or-none propagation (Giugliano et al., 2004), as well as recently for up-state generation in deep layers of cortical slices (e.g., (Capone et al., 2017)). They have been recently revived within the framework of self-organized bi-stability (Scarpetta and de Candia, 2013; di Santo et al., 2016; Scarpetta et al., 2018; Buendía et al., 2020) or quasi-criticality as well (Kinouchi et al., 2020).

We note that spatial subsampling of activity, by recording only from a small number of neurons from the full network, is a common

*Continued Fig. 14 legend from previous page*:

cultures. [Reproduced under CC-NY license from (Levina and Priesemann, 2017).] **D**, Bimodal size distributions in dissociated cultures around the mean interspike interval *Δt* exhibit an initial slope of -2. The mean interspike interval *Δt* has been used as a proxy at which the size distribution should show a power law with slope α = -3/2 (see Fig. 5). [Reproduced with permission from (Yada et al., 2017).] **E**, Bimodal size distribution and linear mean-size-to-duration relation at *Δt* close to the mean interspike interval for dissociated cultures. [Reproduced under CC-NY license from (Shaukat and Thivierge, 2016).] **F**, Spike avalanches of unknown layer origin in organotypic cortex cultures. *Left*: The bimodal size distribution and steepening initial slope with increase in *Δt* suggest hyperexcitable culture condition. *Right*: Near linear mean-size-vs.-duration scaling similar to spike avalanches from dissociated cultures in *E* suggest deviation from critical branching which predicts a slope of 2. [Reproduced under CC-BY license from (Friedman et al., 2012).] **G**, Duration-to-mean-size slope close to 1/2 in dissociated cultures prepared from *postnatal* tissue in line with prediction for a critical branching process. [Reproduced under CC-BY license from (Yaghoubi et al., 2018).] Subpanels *A* – *F* have been modified by adding a *red arrow* emphasizing the bimodal feature in each size distribution and/or a broken red line with slope of -2 as a guide to the eye. A broken red line with slope of -0.5 was added to subpanel *G*.

technical challenge in avalanche analysis. Yet, this technical constraint cannot explain the uncovering of a bimodal distribution at large $\Delta t$. Spatial subsampling decorrelates activity, leading to exponential distributions in cascade sizes (Ribeiro et al., 2014). However, in the present cases, a power law-like or exponential distribution is observed at the outset for spike avalanches at small $\Delta t$, which changes to a bimodal distribution with increasing $\Delta t$ (Fig. 14A – D). In fact, avalanche analysis under increased spatial sampling in dissociated cultures, e.g., using intracellular calcium imaging, established clear bimodal size distribution (Orlandi et al., 2013). Further support that spike avalanches in dissociated cultures differ from LPF avalanches *in vivo* comes from the mean size vs. duration scaling exponent. This exponent was found to be 2, which is in line with expectations for a critical branching process (Miller et al., 2019), but ranges between 1 – 1.5 for spike-based avalanches with bimodal distributions even at large $\Delta t$ (Shaukat and Thivierge, 2016; Yada et al., 2017). This is more in line with expectations for a noise process wherein size simply grows more linearly with duration (Touboul and Destexhe, 2017; Villegas et al., 2019).

We note that in a study by Friedman et al. (2012) spike avalanche distributions were calculated from ten cortex slice cultures and this study is often used as an introduction of scaling relationships for spike avalanches. Three of those cultures exhibited bimodal, four exponential and two were reported as ‘critical’, i.e. power law-like in their size distribution. However, like spike avalanches in dissociated cultures, power law distributions in their cultures considered ‘critical’ steepened in initial slope and became bimodal when increasing $\Delta t$ (Fig. 14F, *left*). The loss of the power law at low temporal resolution supports the interpretation of this activity to be of a 1st order phase transition either from preferentially recording spikes from deep layers or from networks that are hyperexcitable, i.e. ‘supercritical’. This interpretation is further supported by their report of a mean size vs. duration exponent close to 1 (Fig. 14F, *right*; (Friedman et al., 2012)).

### Developmental differences between organotypic cultures and dissociated cultures of cortex

Organotypic cortex cultures that are grown from postnatal brains demonstrate up-states and nested θ/γ–oscillations in their superficial layers, which give rise to avalanche scaling (see Figs. 2 – 7). The conspicuous absence of these dynamics in dissociated cultures suggest an incomplete maturation of superficial layer circuitry, which is supported by several arguments, with the most obvious one being that the standard protocol for dissociated cortex cultures biases towards the formation of deep layer circuits. Dissociated cultures are typically prepared from cortex at embryonic day E18 (Pasquale et al., 2008), which is dominated by deep layer neurons known to autonomously generate population burst activity, also called ‘delta’ brushes (Khazipov and Luhmann, 2006). In contrast, superficial precursor neurons develop relatively late (Luhmann et al., 2016; Molnár et al., 2020), and at E15 – 16 are still migrating towards the cortex along the periventricular wall, a developmental feature that can be used to selectively transfect superficial cells at that developmental stage (Saito, 2006; Bellay et al., 2015) (*cf.* Fig. 13). In addition, late migrating interneurons will be absent in dissociated cultures prepared from the embryonic cortical mantle only. Without endogenous neurotransmitter such as acetylcholine, which is lacking *in vitro*, deep layer pyramidal neurons exhibit intrinsic bursting (McCormick and Prince, 1987; Wang and McCormick, 1993; Compte et al., 2003) that can result in network-wide events (Sanchez-Vives and McCormick, 2000; Wester and Contreras, 2012; Capone et al., 2017). Importantly, organotypic cortex cultures are typically prepared from cortex after birth between postnatal day P1 – 2, at which point most precursor neurons required for establishing superficial pyramidal and interneurons have already arrived in cortex, allowing autonomous assembly of superficial layers in the isolated local cortical culture (see above). This sensitivity to the developmental timepoint of neuronal harvest is further exemplified by cultures of the hippocampus, an evolutionary early part of cortex. Dissociated hippocampus cultures, when taken at E18, reveal avalanche size distribution slope close to -2 and a supercritical branching parameter at 3 ms bin width (Pu et al., 2013). In contrast, dissociated hippocampus cultures made from newborn pups reveal mean size vs. duration scaling exponents of 2, not found for supercritical dynamics (Fig. 14G; (Yaghoubi et al., 2018)). Similarly, Tetzlaff et al. (2010) prepared dissociated cultures from postnatal day P1 – 2 resulting in relatively mild bimodality with increasing bin width (Fig. 14B). Preparing dissociated cultures from postnatal tissue, expansion towards three-dimensional scaffolding using microbeads, and co-culturing with other brain regions, e.g., the hippocampus, might introduce structural heterogeneity that stabilizes avalanche dynamics in future analysis (Massobrio et al., 2015; Brofiga et al., 2020).

To summarize, most avalanche analyses in dissociated cortex cultures reveal power laws that change to a bimodal distribution with *steepening* initial slope at longer integration windows. This dependency on the integration time window seems to reflect a 1$^{st}$ order phase transition commonly found for predominantly deep layer pyramidal networks. These findings suggest that activity in dissociated cultures do not compare well with neuronal avalanche dynamics originally described in organotypic cortex cultures, acute cortex slices and further established in awake *in vivo* preparation that feature neuronal activity localized to superficial layers of cortex.

## Summary and Conclusion

Experimental evidence for SOC in the brain points to the presence of at least four dynamical motifs – up-states, nested oscillations, neuronal avalanches, and coherence potentials. These motifs have been robustly reported for the intact brain and in isolated mammalian cortex, with its layered structure and cell type diversity largely preserved specifically for the organotypic cortex culture and in the acute cortex slice. The co-emergence of scale-invariant neuronal avalanches with oscillations during up-states should encourage future work on SOC in the brain at a disorder-synchronization phase-transition. Neuronal population activity measured in dissociated cortex cultures typically differs from that reported for layered cortex preparations and is more in line with supercritical dynamics. Identifying the precise structural and dynamical constraints responsible for these differences might provide important insights into the mechanisms supporting SOC in the brain.

## Conflict of Interest

The authors declare that the research was conducted in the absence of any commercial or financial relationships that could be construed as a potential conflict of interest.

## Author Contributions

DP took the lead in writing the manuscript and in discussions with all other authors.

## Funding

This research was supported by the Division of the Intramural Research Program of the National Institute of Mental Health (NIMH), USA, ZIAMH002797. This research utilized the computational resources of Biowulf (http://hpc.nih.gov) at the National Institutes of Health (NIH), USA.

## Acknowledgments

We thank Drs. Dante R. Chialvo and Patrick Kanold for comments and support during writing of this manuscript.